\documentclass[journal,onecolumn,a4paper]{IEEEtran}

\usepackage[hidelinks]{hyperref}  

\usepackage{setspace}
\usepackage{times}
\usepackage{mathtools}

\usepackage{amssymb}
\usepackage{amsthm}
\usepackage{amsxtra}
\usepackage{amsmath}
\usepackage{array}
\usepackage{algpseudocode}
\usepackage{algorithm}

\usepackage{verbatim}
\usepackage{epsfig}
\usepackage{setspace}
\usepackage{caption}
\usepackage{subcaption}
\usepackage{breqn}
\usepackage{color}
\usepackage{enumerate}
\usepackage{graphicx}
\usepackage{cite}
\usepackage{microtype}

\allowdisplaybreaks

\newtheorem{theorem}{Theorem}
\newtheorem{definition}{Definition}
\newtheorem{lemma}{Lemma}

\newtheorem{corollary}[theorem]{Corollary}

\newtheorem{claim}{Claim}
\newtheorem{remark}{Remarks}
\newtheorem*{remark*}{Remarks}

\DeclareMathOperator*{\argmin}{argmin}

\newcommand{\cT}{\mathcal{T}}

\newcommand{\BK}{\text{BK}}
\newcommand{\parse}{\text{PARSE}}

\newcommand{\pr}{\text{Pr}}

\newcommand{\rp}{\text{RP}}

\newcommand{\bk}{\textsf{bk}}
\newcommand{\hte}{\textsf{hte}}
\newcommand{\pe}{\textsf{pe}}
\newcommand{\hbk}{\textsf{hbk}}

\newcommand{\E}{\text{E}}

\newcommand{\BSC}{\text{BSC}}

\newcommand{\RC}{\text{RC}}
\newcommand{\cpath}{\text{PATH}}

\newcommand{\ex}{\mathbb{E}}

\newcommand{\bp}{\mathbf{p}}
\newcommand{\ECC}{\text{ECC}}
\newcommand{\trec}{\text{T}^{(\text{r})}}
\newcommand{\tsent}{\text{T}^{(\text{s})}}
\newcommand{\B}{\text{B}}
\newcommand{\X}{\text{X}}
\newcommand{\cc}{\text{c}}
\newcommand{\nc}{\text{nc}}
\newcommand{\NxtLvls}{\text{NxtLvls}}

\algdef{SE}[SUBALG]{Indent}{EndIndent}{}{\algorithmicend }%
\algtext*{Indent}
\algtext*{EndIndent}

\title{Multiparty Interactive Coding over Networks of Intersecting Broadcast Links}

\author{
\IEEEauthorblockN{Manuj Mukherjee$^\dag$} \hspace{1cm} \and \IEEEauthorblockN{Ran Gelles$^\dag$}
}

\begin{document}

\maketitle
\renewcommand{\thefootnote}{}
\footnotetext{
\noindent $^\dag$M.\ Mukherjee and R.\ Gelles are with the Alexander Kofkin Faculty of Engineering, Bar Ilan University, Ramat Gan, Israel. Email: \{mukherm,ran.gelles\}@biu.ac.il. This work was supported in part by the Israel Science Foundation (ISF) through Grant 1078/17.
}
\renewcommand{\thefootnote}{\arabic{footnote}}

\begin{abstract}
We consider computations over networks with multiple broadcast channels that intersect at a single party. Each broadcast link suffers from random bit-flip noise that affects the receivers independently. We design interactive coding schemes that successfully perform any computation over these noisy networks and strive to reduce their communication overhead with respect to the original (noiseless) computation.

A simple variant of a coding scheme by Rajagopalan and Schulman (STOC 1994) shows that any (noiseless) protocol of $R$ rounds can be reliably simulated in  $O(R\log n)$ rounds over a network with $n=n_1n_2+1$ parties in which a single party is connected to $n_2$ noisy broadcast channels, each of which connects  $n_1$ distinct parties. We design new coding schemes with improved overheads. Our approach divides the network into four regimes according to the relationship between $n_1$ and $n_2$. We employ a two-layer coding where the inner code protects each broadcast channel and is tailored to the specific conditions of the regime in consideration. The outer layer protects the computation in the network and is generally based on the scheme of Rajagopalan and Schulman, adapted to the case of broadcast channels.  The overhead we obtain ranges from $O(\log\log n_2)$ to $O(\log n_2 \frac{\log\log n_1}{\log n_1})$ and  beats the trivial $O(\log n)$ overhead in all four regimes.
\end{abstract}

\section{Introduction}\label{sec:intro}

The central problem in communication is ensuring reliable transmission in presence of noise. The pioneering work of Shannon \cite{Shannon48} showed that reliable \emph{one-way communication} is possible in presence of noise. The key idea is to encode the message to be transmitted using an error correcting code. Shannon proved the existence of error correcting codes of positive rate, which ensure reliable one-way communication over a noisy channel. 

One-way communication is not necessarily optimal in terms of communication cost. In a seminal work, Yao \cite{Yao79} considered the scenario where two parties connected by a noiseless channel, and having inputs $x$ and $y$ respectively, wish to compute some function $f(x,y)$. Yao showed that for certain functions $f$, a \emph{two-way interactive communication} protocol can reduce the overall communication cost, as compared to the naive scheme where one party transmits its input to the other party, which subsequently computes the function and sends the output back to the first party. How does one protect the two-way communication protocol if the noiseless link between the parties is replaced by a noisy one? Unfortunately, the naive strategy of protecting each of the messages transmitted during the protocol using an error correcting code fails (see \cite[Section~1.1]{ranbook}). A natural question then is how does one ensure reliable two-way interactive communication in presence of noise. 

The pioneering works of Schulman~\cite{S92,S96} showed that two-way interactive protocols can be reliably simulated in presence of noise with only a constant blow-up in the total amount of communication. The key ingredients in Schulman's simulation were either a \emph{rewind-if-error} mechanism~\cite{S92}, or a two-way equivalent of error correcting codes called \emph{tree codes}~\cite{S96}. Since then, a large body of work has studied the problem of reliable two-way interactive communication, focusing on its various aspects, such as, adversarial noise~\cite{BR14,BE17,GH14,gil}, two-way interactive coding capacity~\cite{KR13,H14,Shay,GHKRW}, computationally efficient encoding and decoding~\cite{Bra12,GMS14,BKN14,GH14}, to mention a few. Reliable simulation of multiparty interactive protocols in presence of noise was first studied by Schulman and Rajagopalan \cite{RS94} for random noise. Later works have examined the minimal overhead of coding schemes for random noise~\cite{BEGH18,ABEGH19,GK19} as well as the scenario with adversarial noise \cite{JKL,HS,CGH19,GKR19}.

The multiparty interactive communication setting studied in \cite{RS94} is described by a graph $G=(V,E)$, where the set of vertices $V$ correspond to the set of $n$ parties, i.e., $|V|=n$, and the set of edges $E$ correspond to the set of point-to-point communication links between pairs of parties. The communication is \emph{fully utilized}, i.e., in every round all parties must send a message to all its neighbours. The communication being interactive, the messages sent in a particular round depends on what was observed in the previous rounds. Schulman and Rajagopalan showed that any fully utilized interactive protocol of $R$ rounds over the graph $G$ can be simulated in presence of random noise in $O(R\log d)$ rounds, where~$d$ is the maximal degree of~$G$. 
The $\log d$ blowup in the number of rounds is not necessary for some graphs, such as cliques \cite{ABEGH19}. On the other hand, super constant blowup is needed for some graphs such as stars~\cite{BEGH18} and cycles~\cite{GK19}.
In this work we wish to extend the question of overhead in multiparty networks to a more general type of networks, namely, networks with \emph{broadcast channels} rather than point-to-point channels.

\subsection{Problem statement}\label{sec:prob}

\begin{figure}[ht]
\centering
\resizebox{0.3\textwidth}{!}{\includegraphics{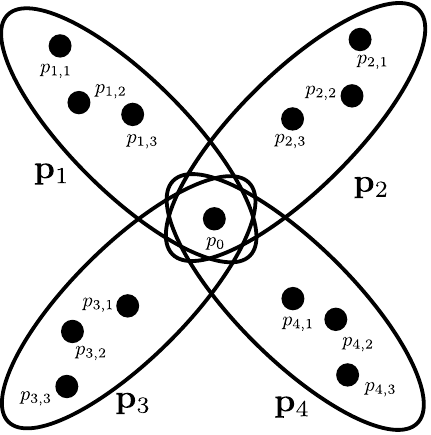}}
\caption{Intersecting broadcast links with $n_1=3$ and $n_2=4$.}
\label{fig:setup}
\end{figure}

For any $n\in\mathbb{N}$, we shall use the notation $[n]\triangleq\{1,2,\ldots,n\}$. We consider the problem of multiparty interactive communication with (single-point) \emph{intersecting} broadcast links. Specifically, we consider networks with $n=n_1n_2+1$ parties that are connected as follows. 
There is one central party~$p_0$, with input $x_0$, and there are $n_2$ \emph{groups} of parties, where each such group consists of $n_1$ distinct parties. For the $i$-th group, we label the parties in that group as $p_{i,j}, j\in[n_1]$, with $x_{i,j}$ being their respective inputs.
We denote the set of all the parties in the $i$-th group as $\bp_i=\{p_{i,j}:j\in[n_1]\}$. An example setup with $n_2=4$ and $n_1=3$ appears in Figure~\ref{fig:setup}. For the ease of notation, we will sometimes use $p_{i,0}$ to denote the central party $p_0$ while describing the communication between $\bp_i$ and $p_0$. Also, we shall use the notation $\bp_{i+0}\triangleq\bp_i\cup\{p_0\}$.

\subsubsection{Communication Model}
We assume each set $\bp_{i+0}$ is connected via a \emph{broadcast channel}.
Any party $p_{i,j}\in\bp_i$ can broadcast bits to all the other parties $p_{i,k}\in\bp_i$, $k\neq j$, and this transmission is also heard by the central party $p_0$. The central party~$p_0$, on the other hand, is connected to the~$n_2$ different broadcast links, and can broadcast to all the parties in $\bp_i$, for every $i\in[n_2]$. We emphasize that a broadcast from $p_0$ meant for parties in $\bp_i$ is not heard by parties in $\bp_j$, $j\neq i$. 

\subsubsection{Fully-utilized (noiseless) protocols}\label{sec:noiseless}
The parties execute a synchronous \emph{protocol}~$\pi$ that determines each party's next bit to broadcast as a function of the party's input and received bits.
The protocol~$\pi$ proceeds in \emph{fully utilized} rounds, i.e., every party must broadcast a single bit in every round. In more detail, in round~$r$, the party $p_{i,j}$ computes one bit $b_{i,j}(r)=\pi(x_{i,j},m_{i,j}^{r-1})$ 
based on its input $x_{i,j}$ and the transcript $m_{i,j}^{r-1}\in\{0,1\}^{n_1\cdot (r-1)}$ it observed so far, and broadcasts this bit over the broadcast channel connecting~$\bp_{i+0}$, i.e., to the remaining parties of $\bp_i$ as well as the central party $p_0$. 
On the other hand, the central party $p_0$, based on its observation $x_0$ and the transcript $m_0^{r-1}\in\{0,1\}^{n_1n_2\cdot(r-1)}$ observed so far, computes a string of $n_2$ bits $b_0(r)=b_{0|1}(r)\ldots b_{0|{n_2}}(r)$, given by $b_{0}(r)=\pi(x_0,m_0^{r-1})$, and broadcasts the $i$-th bit $b_{0|i}(r)$
over the channel connecting $\bp_{i+0}$, for each~$i\in[n_2]$. In the absence of noise, the protocol $\pi$ runs for a fixed number of rounds, the number being independent of the inputs to different parties. This number is referred to as its \emph{round complexity}, and is denoted by $\RC(\pi)$. 
At the end of the $\RC(\pi)$ rounds, each party $p_{i,j}$ outputs its received transcript $\trec_{i,j}\triangleq m_{i,j}^{\RC(\pi)}$. 
Similarly, the central party $p_0$ outputs its received transcript $\trec_0\triangleq m_0^{\RC(\pi)}$. We shall refer to bits sent by the parties $p_{i,j}$ and $p_0$ over the $\RC(\pi)$ rounds as their sent transcripts, and denote them by $\tsent_{i,j}\triangleq b_{i,j}^{\RC(\pi)}$ and $\tsent_0\triangleq b_0^{\RC(\pi)}$, respectively.

\subsubsection{Noise model}\label{sec:noise}
We consider the scenario where every broadcast link suffers independent random bit flips. 
Namely, for a bit $b$ broadcast by $p_{i,j}$ at round~$r$, the party $p_{i,k}, 0\leq k\leq n_1, k\neq j,$ receives the $\epsilon$-noisy copy $b\oplus \mathsf{err}$, where $\mathsf{err}\sim \text{Ber}(\epsilon)$ is independent across the senders $p_{i,j}$, the receivers $p_{i,k}$, as well as the across different rounds $r$. 

We call a simulation $\Pi$ of a noiseless protocol $\pi$ successful, if at the end of $\Pi$ the parties $p_{i,j}$ (resp., $p_0$) output $\tsent_{i,j},\trec_{i,j}$ (resp., $\tsent_0, \trec_0$).\footnote{Note that in the presence of noise, the parties do not broadcast the correct bits all the time. Hence, estimates of both received and sent transcript need to be calculated by the parties.}

\subsection{Our results}\label{sec:res}

Successful simulation of interactive communication has been studied for two extreme cases of this model. 
The first case, with $n_1=1$, and hence $n=n_2+1$, reduces to a star network with point-to-point links, studied by Braverman et al.~\cite{BEGH18}. It has been shown in~\cite{BEGH18} that any protocol of $R$ rounds can be simulated in $O(R\frac{\log n}{\log\log n})$ rounds in presence of random noise. 
The second extreme case, with $n_2=1$, and hence $n=n_1+1$, reduces to a single broadcast link, studied by Gallager~\cite{Gal}. 
In this case, it is known that exchanging a single bit between the parties requires $\Theta(\log\log n)$ rounds \cite{Gal,GKS} in presence of random noise. In this paper, we extend the above two extremes and examine the family of intersecting broadcast links with arbitrary $n_1$ and $n_2$. 

Interestingly, the overhead of interactive coding in such networks is sensitive to the relations between $n_1$ and $n_2$. We will assume both $n_1,n_2$ tend to infinity and divide the universe of possible networks into four regimes capturing four different asymptotic constraints on $n_1$ and $n_2$. For each such regime we design a coding scheme that is tailored to that sub-family of networks, and obtain a scheme with non-trivial overhead.
Our results are captured by the following theorem.

\begin{theorem}\label{th:main}
Let~$\pi$ be a fully utilized, noiseless protocol with $R$~rounds, designed to run on a network of intersecting broadcast links consisting of $n=n_1n_2+1$ parties. 
Then, there exists a fully utilized simulation~$\Pi$ of the original protocol $\pi$ over $\epsilon$-noisy broadcast channels.
The simulation~$\Pi$ runs for $O(R\cdot h_{n_1,n_2})$ rounds and succeeds except with a vanishing probability of error. 

The quantity $h_{n_1,n_2}$ is given by
\begin{equation*}
h_{n_1,n_2}=
\begin{cases}
\log\log n_1, & \text{if }n_2=2^{O(\log n_1)}\\
\log\log n_2, & \text{if }n_2=2^{\Omega(\log n_1)} \text{ but } n_2=2^{O(n_1)}\\
\frac{\log n_2\log\log n_1}{\log n_1}, & \text{if }n_2=2^{\Omega(n_1)} \text{ but }n_2=2^{2^{O(\frac{n_1\log n_1}{\log\log n_1})}}\\
\frac{n_1\log n_2}{\log\log n_2}, & \text{if }n_2=2^{2^{\Omega(\frac{n_1\log n_1}{\log\log n_1})}}.
\end{cases}
\end{equation*}
\end{theorem}

As noted previously,  Gallagher~\cite{Gal} gave a scheme for $n$ parties connected by a single noisy broadcast link (i.e., $n_2=1$), that allow the parties to successfully exchange a single bit with high probability, using $O(\log\log n)$ fully utilized communication rounds. As a corollary to our result in Theorem~\ref{th:main}, we can extend the result of \cite{Gal} to show that any fully-utilized noiseless protocol $\pi$ over the (single-hop) broadcast link, with $R$~rounds, can be successfully simulated in presence of noise in $O(R\cdot\log\log n)$ rounds with high probability.
\begin{corollary}
\label{cor:broadcast}
Let~$\pi$ be a fully utilized, noiseless protocol with $R=\RC(\pi)$ rounds, designed to run on a single broadcast link with $n$ parties. 
Then, there exists a fully utilized simulation $\Pi$ of the original protocol $\pi$ over $\epsilon$-noisy broadcast channel. The simulation~$\Pi$ runs for $O(R\cdot \log\log n)$ rounds and succeeds except with a vanishing probability of error. 
\end{corollary}

\begin{IEEEproof}
The result follows by plugging in $n_2=1$ and $n_1=n-1$ in Theorem~\ref{th:main}.\footnote{This is valid even though $n_2$ is not growing but a constant. This is because in the proof for the first regime (see Section~\ref{sec:reg1}), we only need $n_1$ to be growing in order to decrease the probability of error.}
\end{IEEEproof}

\subsection{Techniques overview}\label{sec:pt}

Our simulation consists of two layers of coding. The inner layer protects the communication in each broadcast channel. This layer guarantees that $p_0$ hears most of the broadcast bits with high probability; the coding itself depends on the relation between $n_1$ and $n_2$. 
The outer layer protects the communication with respect to~$p_0$, that is, making sure that even if $p_0$ obtains an incorrect bit in some broadcast channel (despite the inner code), this will cause a bounded delay in the progress of the protocol throughout the entire network. This layer is based on the coding  technique by Rajagopalan and Schulman~\cite{RS94} adapted to the case of broadcast channels.

The simulation $\Pi$ is designed as follows. We first show how to adapt the Rajagopalan-Schulman (RS) scheme to the case where parties communicate over broadcast links rather than  point-to-point links. This immediately gives us the result that a simulation $\Pi$ of $O(R\log n)$ rounds exists, for any protocol $\pi$, irrespective of the relationship between $n_1$ and $n_2$. However, we are able to reduce the number of rounds by further modifying the RS scheme. Consider the first regime, $n_2=2^{O(\log n_1)}$, where the number of broadcast links is at most comparable to the size of each link. The RS scheme involves broadcasting tree coded symbols in each round. However, instead of directly broadcasting these tree coded symbols, for this regime, we shall have the parties exchange them using the bit exchange scheme of~\cite{Gal}. Since $n_2$ is relatively small in this regime, the overhead will be dominated by the $O(\log \log n_1)$ term required to exchange the bits in every broadcast channel. However, in the other regimes, the overhead of $O(\log n_2)$ imposed by the outer layer becomes more and more dominant and we need to carefully tailor the inner code to obtain superior overhead.

In more details, for the second regime the parties first exchange their bits in small groups and then use error correcting code to amplify the probability that $p_0$ correctly obtains the bits of all (sub-)groups.
In the third regime, a simple repetition of Gallager's scheme allows amplifying the success probability while avoiding the overhead of~$O(\log n)$. 
In the final regime, 
the graph begins to look like the star graph, with the number of broadcast links being much larger than the size of each group. 
Accordingly, we employ a suitable variant of the scheme of~\cite{BEGH18},  which was originally designed for the point-to-point star graph. 
The key idea here is to simulates multiple rounds of the original protocol at one go; namely, the parties collect the communication of $k$ consecutive rounds of~$\pi$ and employ error-correction codes on the aggregated word. 
Note that due to the interactive nature of the problem, parties \emph{do not know in advance} the messages they will send in~$\pi$, as those depend on the incoming messages of future rounds. Instead, the parties communicate all the possible options for the next~$k$ rounds. The central party~$p_0$ collects all these transmissions and computes the progress $\pi$ makes in these $k$~rounds. 
Note, however, that sending all the possible options incurs a large overhead, thus the value of $k$ must be small; at the same time a larger value of $k$ will allow better aggregation, which reduces the overhead. Optimizing this tradeoff leads to the overhead in our last regime.

\subsection{Notation and conventions}\label{sec:not}

The description of the simulation $\Pi$ and its proof involves a lot of notation. In this section, we give a summary of the notation convention used throughout the paper. The conventions are followed unless explicitly stated. For any $m\in\mathbb{N}$, we shall use the notation $[m]\triangleq\{1,2,\ldots,m\}$. We shall use $\mathbf{1}\{\cdot\}$ to denote the indicator function. Since our result involves asymptotic regimes only, for the sake of brevity, we shall drop clarifications like `for all sufficiently large values of $n$.' For example, we may state the relation $5n_1\leq n_1^2$ without clarification, even though it is only valid for $n_1\geq 5$.

Alphabets of codes are denoted by upper case letters or capitalized Greek letters. Given an alphabet $\Lambda$, the alphabet for strings from $\Lambda$ of length $l$ will be denoted by $\Lambda^l\triangleq\underbrace{\Lambda\times\ldots\times\Lambda}_{l\text{ times}}$. On the other hand, $\Lambda^*$ will be used to denote the alphabet of $\Lambda$-valued strings of arbitrary length. Strings, including strings of vector valued symbols, are denoted by lower case letters, or lower case Greek letters. We shall use $|\cdot|$ to denote the function which gives the length of a string. Given any string $\theta$ (can be vector valued) of length $l$, we shall denote the $i$th symbol in that string using $\theta(i)$, and denote any $k\leq l$ length substring of $\theta$ by $\theta^k\triangleq\theta(1)\ldots\theta(k)$. Also, a substring of $\theta$ from its $k_1$th to $k_2$th location, $1\leq k_1<k_2\leq l$, will be denoted by $\theta(k_1:k_2)\triangleq \theta(k_1)\ldots\theta(k_2)$.

In our simulation protocol, generally a string related to a certain party will have its identity as subscript. For example, a string $a_{i,j}$ is associated with party $p_{i,j}$. Furthermore, the symbols of $a_{i,j}$ may themselves be vectors of $n_1$ components, each component relating to a party $p_{i,j'}\in\bp_{i+0}$, with $j'\neq j$. In such cases, we shall denote the $l$th symbol of $a_{i,j}$ as $a_{i,j}(l)\triangleq (a_{i,j|j'}(l))_{j'\neq j}$, where $a_{i,j|j'}(l)$ corresponds to the component of $a_{i,j}(l)$ related to the party $p_{i,j'}$. Going by previously mentioned convention, we shall have $a_{i,j|j'}^l\triangleq a_{i,j|j'}(1)\ldots a_{i,j|j'}(l)$. Similarly, strings associated with party $p_0$, where each symbol is constituted of individual components relating to parties $p_{i,j}$ (resp. broadcast links $\bp_i$), shall be denoted as $a_0(l)=(a_{0|i,j}(l))_{i\in[n_2],j\in[n_1]}$ (resp. $a_0(l)=a_{0|1}(l)\cdots a_{0|n_2}(l)$). During the runs of the simulation $\Pi$, we will encounter strings whose lengths get appended by one symbol per step. Hence, as per our convention, a simple superscript of $r$ is enough to denote such strings after $r$ steps. However, there will also be strings which are updated by both appending as well as deleting symbols. For such strings, we shall be using $a^{(r)}$ to denote realization of the string $a$ at step $r$. If $|a^{(r)}|=l$, then any $k\leq l$ length substring of $a^{(r)}$ will be denoted by $a^{(r)l}$. Finally, to avoid confusion between $l$ length substrings of $a^{(r-1)}$, and a substring of $a$ of length $(r-1)\cdot l$, we shall respectively use the notations $a^{(r-1)l}$ and $a^{(r-1)\cdot l}$. 

\subsection{Organization}\label{sec:org}

Technical building blocks like error correcting codes, tree codes, concentration inequalities are introduced in the following Section~\ref{sec:tools}. Section~\ref{sec:RS} introduces the RS simulation with broadcast links. Simulation schemes for the first three regimes of Theorem~\ref{th:main} are described in Section~\ref{sec:bitex}. The simulation $\Pi$ for the final regime is given in Section~\ref{sec:star}. Proofs of a few technical results are relegated to the appendices, which are available as supplementary material to this document.

\section{Some basic tools}\label{sec:tools}

\subsection{Codes}\label{sec:ecc}

The first tool we need is the standard error correcting code whose properties are described in the following lemma.

\begin{lemma}[Theorem~5.6.2 of \cite{ITRC}]\label{lem:ecc}
Consider a binary symmetric channel with capacity $C$. For any $k,m$ where $m=\Theta(\frac{k}{C})$, there exists an error correcting code $\ECC:\{0,1\}^k\to\{0,1\}^m$ with probability of decoding error over the channel being $2^{-\Omega(m)}$.
\end{lemma}

The following lemma showing the existence of an error correcting code with relative distance arbitrarily close to half, but rates constant in the input length, is essentially a restatement of the \emph{Gilbert Varshamov bound} (see Theorem~5.1.7 of \cite{vanlint}). 

\begin{lemma}\label{lem:KGDR}
Let $\delta>0$, and fix $K=\lceil\frac{10}{\delta^2}\rceil$. Then, for any $m>0$, there exists an error correcting code $\ECC_\delta:\{0,1\}^m\to\{0,1\}^{mK}$ satisfying the following: For any $s,t\in\{0,1\}^m$ such that $s\neq t$, we have $\Delta(\ECC_\delta(s),\ECC_\delta(t))>(\frac{1}{2}-\delta)Km$.
\end{lemma}
A proof of Lemma~\ref{lem:KGDR} in its present form is available in \cite[Lemma~4.6]{KGDR}.

In order to secure our simulated protocol from noise, we will also require the combinatorial object referred to as a tree code. A tree code $\cT:[d]^*\to S^*$ with distance $\alpha\in(0,1)$ is defined as follows. Consider a complete $d$-ary tree where the $d$ outgoing edges of a node are labeled as $1,2,\ldots,d$. Thus, any string in $[d]^l$, for every $l>0$, corresponds to a unique path of depth $l$ in the tree. The tree code $\cT:[d]^*\to S^*$ associates a symbol from $S$ to every edge of the $d$-ary tree. Thus, for any string $w\in[d]^l$, $\cT(w)$ is obtained by collecting the $l$ symbols from $S$ associated with the edges in the depth-$l$ path corresponding to $w$, starting from the root. Moreover, for any $w,z\in[d]^l$, whose least common ancestor is at some level $h\geq 0$, $\cT$ satisfies $\Delta(\cT(w),\cT(z))\geq\alpha(l-h)$.

The following lemma from \cite{S96} shows that tree codes indeed exist. 

\begin{lemma}[Lemma~1 of \cite{S96}]\label{lem:ex}
For any $\alpha\in(0,1)$, and any positive integer $d$, there exists a tree code $\cT$ of distance $\alpha$ on $d$-ary trees, with output alphabet $S$ satisfying
$$
|S|=2\lfloor(2^{h(\alpha)}2d)^{\frac{1}{1-\alpha}}\rfloor-1,
$$ 
where $h(p)\triangleq-p\log_2(p)-(1-p)\log_2(1-p)$ is the binary entropy function. 
\end{lemma}

\subsection{Probability bounds}\label{sec:bound}

We shall employ the following version of the \emph{Chernoff bound} to bound probabilities of certain error events. 

\begin{lemma}[Theorem~4.5 of \cite{MU05}]\label{lem:chern}
Consider i.i.d. random variables $X_1,X_2,\ldots,X_m$ with $\ex[\sum_{i=1}^mX_i]=\mu$. Fix any $0\leq \delta\leq 1$. Then the following holds:
$$
\text{Pr}(\sum_{i=1}^mX_i\leq\mu(1-\delta))\leq e^{-\frac{\delta^2\mu}{2}}.
$$
\end{lemma}

Finally, we state the standard bound on the probability of error of decoding a repeated bit across a binary symmetric channel (BSC) using majority decoding.
\begin{lemma}[Section~5.3 of \cite{ITRC}]\label{lem:repeat}
Consider the scenario where the sender sends $\rho$ copies of a bit across a $\BSC$ of crossover probability $\epsilon<\frac{1}{2}$. The receiver decodes the sent bit by the majority rule. Then, the probability of error in decoding, $\text{P}_e$, is bounded as
$$
\text{P}_e\leq (2\sqrt{\epsilon(1-\epsilon)})^\rho.
$$
\end{lemma}

\section{A Rajagopalan--Schulman scheme for intersecting broadcast links}\label{sec:RS}

The Rajagopalan--Schulman (RS) scheme of \cite{RS94} gives a method for simulating an interactive protocol over any network of arbitrary topology having noisy point-to-point links. 
A broadcast channel is similar to having multiple point-to-point channels under the restriction that a party must send the same bit over each one of its adjacent channels, in every round.
On its surface, the RS scheme does not restrict the parties to sending the same message over all its adjacent channels and, thus, is inapplicable over broadcast channels.
What saves us is the fact that $\pi$ itself assumes broadcast channels, hence, as long as no error happens, the RS  merely runs $\pi$ which can work on broadcast channel. 
If an error happens, the RS scheme needs to instruct some of the parties to roll-back and revert their simulation to a prior state preceding the error; other parties might need to ``hold'' their state as their simulation state already precedes the state of other parties. What saves us here is that the actions of a single party~$p_{i,j}$ affects only the parties in~$\bp_i$, to which $p_{i,j}$~can broadcast its own action (i.e., continuing or rewinding). Then, a fully-utilized broadcast channel suffices to allow all the members in~$\bp_{i+0}$ to communicate their continue/rewind instructions. Other sets $\bp_k$ with $k\ne i$ behave independently and are not affected by the actions of the parties in~$\bp_i$.

We now detail the exact variant of the RS scheme we shall use. 
The simulation $\Pi$ uses a ternary alphabet $\Sigma\triangleq\{0,1,\BK\}$. To protect the transmissions in $\Pi$ from channel errors, the parties employ a tree code $\cT:\Sigma^*\to S^*$ of distance $\alpha=0.5$. Note that by Lemma~\ref{lem:ex}, we have $|S|=c'$, for some constant $c'>0$, and hence the symbols from $S$ can be encoded to $c=\lceil\log c'\rceil$ bits. 

We next specify how the parties decode the tree code. Consider the party $p_{i,j}$ trying to decode the bits it has received from the party $p_{i,k}$, $0\leq k\leq n_1, k\neq j$, after $r$ rounds. The party $p_{i,j}$ first decodes all the $rc$ bits it has received so far from $p_{i,k}$ to obtain the vector $y^r\in S^r$. It then obtains an estimate $\hat{w}^r\in\Sigma^r$ of the transmissions so far by $p_{i,k}$, using the following rule:
\begin{equation}
\hat{w}^r=\displaystyle\argmin_{z^r\in\Sigma^r}\Delta(y^r,\cT(z^r)). \label{eq:dec}
\end{equation} 
The decoding at all other parties proceed similarly, including the central party. 

In order to describe the simulation $\Pi$, we first introduce the parse operation. 
\begin{definition}
\label{def:parse}
A string $x\in\Sigma^*$ is called \emph{parseable} if every prefix of the string contains at least as many non-$\BK$ symbols as $\BK$ symbols. The operation $\parse:\Sigma^*\to\{0,1\}^*$ on parseable strings is defined as the removal of every symbol $\BK$ symbol along with the first non-$\BK$ symbol preceding it. If $x\in\Sigma^*$ is not parseable, then set $\parse(x)=\emptyset$.
\end{definition}
For example, $\parse(00\BK11\BK\BK10)=010$, whereas $\parse(1\BK\BK001)=\emptyset$.
\begin{remark}
\label{rem:parse}
It is easy to verify the following:
\begin{enumerate}
\item Any parseable string $x$ with $x'=\parse(x)$ satisfies $|x'|=|x|-2\sum_{i=1}^{|x|}\mathbf{1}\{x(i)=\BK\}$. 
\item If $x'\neq\emptyset$, then the concatenation of $x$ with $\BK$, i.e., $x\circ\BK$, is parseable.
\end{enumerate}
\end{remark}

\begin{algorithm}
\caption{The simulation $\Pi$ (actions taken by the parties $p_{i,j}$, $i\in[n_1],j\in[n_2]$)}
\label{alg:rsperip}
\textbf{input:} 
\begin{enumerate}[{1.}]
\item Protocol $\pi$ with zero-padding after $\RC(\pi)$ transmissions
\item Input $x_{i,j}$
\item A tree-code $\cT:\Sigma^*\to S^*$ with distance $\alpha=\frac{1}{2}$
\end{enumerate}
\begin{algorithmic}[1]
\State \textbf{initialize} $w_{i,j}\leftarrow\emptyset, s_{i,j}\leftarrow\emptyset$
\Statex

\For{$r=1$ \textbf{to} $2\RC(\pi)$}
\State Decode $\hat{w}_{i,j}^{r-1}$ 
\State $u_{i,j}\leftarrow$ ConsCheck$(\pi,(i,j),x_{i,j},\hat{w}_{i,j}^{r-1},w_{i,j}^{r-1},s_{i,j}^{r-2})$ 
\If{$u_{i,j}=1$} \Comment{Consistency checks}                         
\State $s_{i,j|k}(r-1)\leftarrow\hat{w}_{i,j|k}(r-1), \; \; \;  \forall 0\leq k\leq [n_1], k\neq j$
\State $w_{i,j}(r)\leftarrow\pi\biggl(x_{i,j},\sigma_{i,j}^{(r)}\biggr)$
\Else
\State $s_{i,j}(r-1)\leftarrow\BK^{n_1}$, $w_{i,j}(r)\leftarrow\BK$      \Comment{Moving back by one step, if inconsistent}
\EndIf
\State BCast$(\cT(w_{i,j}^r)(r))$ with $\bp_{i+0}$     
\Comment{Use relevant broadcast algorithm to exchange $r$th symbol from $S$ of $\cT(w_{i,j}^r)$ with $\bp_{i+0}$}
\EndFor
\Statex

\State Decode $\hat{w}_{i,j}^{2\RC(\pi)}$ \Comment{Decoding the final transmission from the other parties}
\State $u_{i,j}\leftarrow$ ConsCheck$(\pi,(i,j),x_{i,j},\hat{w}_{i,j}^{2\RC(\pi)},w_{i,j}^{2\RC(\pi)},s_{i,j}^{2\RC(\pi)-1})$
\If{$u_{i,j}=1$}
\State $s_{i,j|k}(2\RC(\pi))\leftarrow\hat{w}_{i,j|k}(2\RC(\pi)), \;   \forall 0\leq k\leq [n_1], k\neq j$, \;\; $w_{i,j}(2\RC(\pi)+1)\leftarrow 0$
\Else
\State $s_{i,j}(2\RC(\pi))\leftarrow\BK^{n_1}$, $w_{i,j}(2\RC(\pi)+1)\leftarrow\BK$
\EndIf
\State $\hat{\text{T}}^{(\text{s})}_{i,j}\leftarrow\parse(w_{i,j}^{2\RC(\pi)+1})$, $\hat{\text{T}}^{(\text{r})}_{i,j}\leftarrow\biggl(\parse(s_{i,j|k}^{2\RC(\pi)}):0\leq k\leq n_1, k\neq j\biggr)$\Comment{Calculate estimated transcript}
\Statex\Statex

\State\Return $(\hat{\text{T}}^{(\text{s})}_{i,j},\hat{\text{T}}^{(\text{r})}_{i,j})$
\end{algorithmic}
\end{algorithm}

\begin{algorithm}
\caption{The simulation $\Pi$ (actions taken by party $p_0$)}
\label{alg:rscent}
\textbf{input:} 
\begin{enumerate}[{1.}]
\item Protocol $\pi$ with zero-padding after $\RC(\pi)$ transmissions
\item Input $x_0$
\item A tree-code $\cT:\Sigma^*\to S^*$ with distance $\alpha=\frac{1}{2}$
\end{enumerate}
\begin{algorithmic}[1]
\State \textbf{initialize} $w_0\leftarrow\emptyset, s_0\leftarrow\emptyset$
\Statex

\For{$r=1$ \textbf{to} $2\RC(\pi)$}
\State Decode $\hat{w}_0^{r-1}$ 
\State $u_0\leftarrow$ ConsCheck$(\pi,0,x_0,\hat{w}_0^{r-1},w_0^{r-1},s_0^{r-2})$ 
\If{$u_0=1$} \Comment{Consistency checks}                         
\State $s_{0|i,k}(r-1)\leftarrow\hat{w}_{0|i,k}(r-1), \; \; \;  \forall i\in[n_2], k\in[n_1]$
\State $w_0(r)\leftarrow\pi\biggl(x_0,\sigma_0^{(r)}\biggr)$
\Else
\State $s_0(r-1)\leftarrow\BK^{n_1n_2}$, $w_0(r)\leftarrow\BK^{n_2}$      \Comment{Moving back by one step, if inconsistent}
\EndIf
\State BCast$(\cT(w_{0|i}^r)(r))$ with $\bp_i$, $\forall i\in[n_2]$        \Comment{Use relevant broadcast algorithm to exchange $r$th symbol from $S$ of $\cT(w_{0|i}^r)$ with $\bp_{i}$}
\EndFor
\Statex

\State Decode $\hat{w}_0^{2\RC(\pi)}$ \Comment{Decoding the final transmission from the other parties}
\State $u_0\leftarrow$ ConsCheck$(\pi,0,x_0,\hat{w}_0^{2\RC(\pi)},w_0^{2\RC(\pi)},s_0^{2\RC(\pi)-1})$
\If{$u_{i,j}=1$}
\State $s_{0|i,k}(2\RC(\pi))\leftarrow\hat{w}_{0|i,k}(2\RC(\pi)), \;   \forall i\in[n_2], k\in[n_1]$, \;\;$w_{0}(2\RC(\pi)+1)\leftarrow 0^{n_2}$
\Else
\State $s_{0}(2\RC(\pi))\leftarrow\BK^{n_1n_2}$, $w_0(2\RC(\pi)+1)\leftarrow\BK^{n_2}$
\EndIf
\State $\hat{\text{T}}^{(\text{s})}_0\leftarrow\biggl(\parse(w_{0|i}^{2\RC(\pi)+1}):i\in[n_2]\biggr)$, $\hat{\text{T}}^{(\text{r})}_0\leftarrow\biggl(\parse(s_{0|i,k}^{2\RC(\pi)}):i\in[n_2], k\in[n_1]\biggr)$ \Comment{Calculate estimated transcript}
\Statex
\State\Return $(\hat{\text{T}}^{(\text{s})}_0,\hat{\text{T}}^{(\text{r})}_0)$ 
\end{algorithmic}
\end{algorithm}

\begin{algorithm}
\caption{ConsCheck() for step $r$}
\label{alg:cons}
\begin{algorithmic}[1]
\Statex \textit{At party $p_{i,j}, \forall i\in[n_2],j\in[n_1]$:}
\Indent 
\State $u\leftarrow 1$
\State $\hat{z}_{i,j|k}^{(r)}\leftarrow \parse(\hat{w}_{i,j|k}^{r-1}), \forall 0\leq k\leq [n_1], k\neq j$
\State $\hat{\ell}_{i,j}^{(r)}\leftarrow\min_{\substack{0\leq k\leq n_1\\ k\neq j}}|\hat{z}_{i,j|k}^{(r)}|$
\If{$\hat{\ell}_{i,j}^{(r)}\leq \ell_{i,j}^{(r-1)}$}
\State $u\leftarrow 0$
\Else
\For{$l=1$ \textbf{to} $\ell_{i,j}^{(r-1)}$}
\State $u\leftarrow u\wedge\biggl(\wedge_{\substack{0\leq k\leq n_1\\ k\neq j}}\mathbf{1}\biggl\{\hat{z}_{i,j|k}^{(r)}(l)=\sigma_{i,j|k}^{(r-1)}(l)\biggr\}\biggr)\wedge\mathbf{1}\{z_{i,j}^{(r-1)}(l+1)=\pi(x_{i,j},\sigma_{i,j}^{(r-1)l})\}$ 
\Statex \Comment{Checking if estimated transcripts agree with $\pi$}
\EndFor
\EndIf
\State \Return $u$
\EndIndent
\Statex
\Statex \textit{At party $p_0$:}
\Indent 
\State $u\leftarrow 1$
\State $\hat{z}_{0|i,k}^{(r)}\leftarrow \parse(\hat{w}_{0|i,k}^{r-1}), \forall i\in[n_2], k\in[n_1]$
\State $\hat{\ell}_0^{(r)}\leftarrow\min_{i\in[n_2], k\in[n_1]}|\hat{z}_{0|i,k}^{(r)}|$
\If{$\hat{\ell}_0^{(r)}\leq \ell_0^{(r-1)}$}
\State $u\leftarrow 0$
\Else
\For{$l=1$ \textbf{to} $\ell_0^{(r-1)}$}
\State $u\leftarrow u\wedge\biggl(\wedge_{i\in[n_2], k\in[n_1]}\mathbf{1}\biggl\{\hat{z}_{0|i,k}^{(r)}(l)=\sigma_{0|i,k}^{(r-1)}(l)\biggr\}\biggr)\wedge\mathbf{1}\{z_0^{(r-1)}(l+1)=\pi(x_0,\sigma_0^{(r-1)l})\}$ 
\State \Comment{Checking if estimated transcripts agree with $\pi$}
\EndFor
\EndIf
\State \Return $u$
\EndIndent

\end{algorithmic}
\end{algorithm}

The simulation $\Pi$ proceeds in steps. At the end of the $r$-th step, the party $p_{i,j}$ possesses the following strings: $w_{i,j}^r\in\Sigma^r$ which consists of the symbols it has broadcast so far, as well as the strings $s_{i,j}^{r-1}=(s_{i,j|k}^{r-1}:0\leq k\leq n_1, k\neq j)\in\Sigma^{n_1\cdot(r-1)}$, which are its estimates of the strings of the symbols sent by the parties $p_{i,k}, 0\leq k\leq n_1, k\neq j$. Similarly, the central party $p_0$ has the string $w_0^{n_2r}=(w_{0|i}:i\in[n_2])\in\Sigma^{n_2r}$ of symbols it has sent to the broadcast links $\bp_i, i\in[n_2]$, so far, as well as its estimate $s_0^{r-1}=(s_{0|i,k}^{r-1}:i\in[n_2], k\in[n_1])\in\Sigma^{n_1n_2r}$ of the symbols sent by the parties $p_{i,k}, i\in[n_2], k\in[n_1]$.\footnote{A different notation, $s$ and $w$, with suitable subscripts, is used to denote the received and sent symbols in the simulation $\Pi$, as opposed to respectively $m$ and $b$ for the noiseless protocol $\pi$. This is done for a couple of reasons. Firstly, there is a need to distinguish between the messages sent and received in the noiseless case from the noisy case. This is important in the proof where we check how much of the transmitted and received messages in the noisy setting agree with those in the noiseless setting. Secondly, the messages exchanged in simulation $\Pi$ are not bits but come from $\{0,1,\BK\}$, whereas $b$ and $m$ are bits.} Note that since till the $r$th step, all parties only decode $r-1$ transmissions.

We call the parsed versions of the sent and estimated strings available at a party its estimated transcript. The estimated transcript of a party, corresponds to its belief that how much of the protocol $\pi$ it has already simulated. More formally, let $z_{i,j}^{(r)}\triangleq\parse(w_{i,j}^r)$, $z_{0|i}^{(r)}\triangleq\parse(w_{0|i}^r)$, $\sigma_{i,j|k}^{(r)}\triangleq\parse(s_{i,j|k}^{r-1})$, and $\sigma_{0|i,k}^{(r)}\triangleq\parse(s_{0|i,k}^{r-1})$. Then, the estimated transcript of $p_{i,j}$ after step $r$ consists of $z_{i,j}^{(r)}$ and $\sigma_{i,j}^{(r)}=(\sigma_{i,j|k}^{(r)}: 0\leq k\leq [n_1], k\neq j)$. Similarly, the estimated transcript of $p_0$ after step $r$ consists of $z_0^{(r)}=(z_{0|i}^{(r)}:i\in[n_2])$, and $\sigma_0^{(r)}=(\sigma_{0|i,k}^{(r)}:i\in[n_2],k\in[n_1])$. We further note that simulation $\Pi$ (see Algorithm~\ref{alg:rsperip}) ensures that for any party $p_{i,j}$, at the end of any step $r$, if $w_{i,j}(l)=\BK$, then $s_{i,j|k}(l-1)=\BK$. This implies that $|z_{i,j}^{(r)}|-1=|\sigma_{i,j|k}^{(r)}|\triangleq\ell_{i,j}^{(r)}$. Similarly, one can argue that for the central party $p_0$, all estimated sent transcripts $z_{0|i}^{(r)}, i\in[n_2]$, are of the same length, and they satisfy $|z_{0|i}^{(r)}|-1=|\sigma_{0|i,k}^{(r)}|\triangleq\ell_0^{(r)}$.

The actions taken by parties $p_{i,j}$ and $p_0$ during the simulation $\Pi$ are detailed in Algorithms~\ref{alg:rsperip} and~\ref{alg:rscent} respectively. The simulation consists of $T=2\RC(\pi)$ steps. At the beginning of the $r$th step, the party first decodes all the symbols it has received so far from the previous $r-1$ steps. We denote the strings decoded by party $p_{i,j}$ (resp. $p_0$) by $\hat{w}_{i,j}^{r-1}=(\hat{w}_{i,j|k}^{r-1}: 0\leq k\leq [n_1], k\neq j)\in\Sigma^{n_1\cdot(r-1)}$ (resp., $\hat{w}_{0}^{r-1}=(\hat{w}_{0|i,k}^{r-1}: i\in[n_2], k\in[n_1])\in\Sigma^{n_1n_2\cdot(r-1)}$. It next checks, whether these decoded symbols together with its estimated transcripts from the $(r-1)$th step are consistent with the original protocol $\pi$ (see Algorithm~\ref{alg:cons}). If it finds they are indeed consistent, it calculates the next symbol(s) to be sent according to the protocol $\pi$.\footnote{We zero-pad $\pi$ to ensure there is always a next symbol to send.} If however, the party detects an inconsistency between what it decoded and its estimated transcript, it decides to broadcast $\BK$. Once the next symbol is decided, the parties append it to the string of already sent $r-1$ symbols, and applies the tree code $\cT$ on this string. The $r$th symbol of this tree coded string is then exchanged with the other members of its broadcast link (including $p_0$) using a broadcast algorithm BCast$()$. Note that the central node takes part in $n_2$ instances of the broadcast algorithm simultaneously, one instance per broadcast link. 

Different variants of the broadcast algorithm BCast$()$ are used to obtain the results of first three regimes of Theorem~\ref{th:main}. See Section~\ref{sec:bitex} for the details of these broadcast algorithms. The final regime of $n_2=2^{2^{\Omega(\frac{n_1\log n_1}{\log\log n_1})}}$, however, requires a different simulation protocol $\Pi$, which is described in Section~\ref{sec:star}.

The following theorem characterizes the performance of the simulation $\Pi$.
\begin{theorem}
Consider intersecting broadcast links with $n=n_1n_2+1$ parties, affected by random $\epsilon$-bit-flip noise. Suppose that each broadcast symbol from $S$ is received correctly by a recipient, except with probability~$p$. Then, given any noiseless protocol $\pi$, the simulation $\Pi$ given in Algorithms ~\ref{alg:rsperip} and~\ref{alg:rscent} is successful except with probability~$(32n^5p^{\frac{1}{16}})^{\RC(\pi)}$.
\label{th:rs}
\end{theorem}

The proof of Theorem~\ref{th:rs} closely follows the approach in~\cite{rajathes}, and is included in Appendix~\ref{app:rs} in the supplementary material for completeness. 

A trivial example of a broadcast algorithm BCast$()$ is as follows. Choose an error correcting code $\ECC:S\to\{0,1\}^{C\log n}$, for some constant $C>0$, and broadcast the symbol from $S$ after encoding it with $\ECC$. A single broadcast thus takes a total of $C\log n$ rounds, and hence the round complexity of the simulation protocol $\Pi$ is $\RC(\pi)O(\log n)$. Further, one can claim similar to Lemma~\ref{lem:eccex} in Section~\ref{sec:star}, that the constant $C$ maybe chosen large enough (independent of $n$) to ensure that any broadcast symbol from $S$ is decoded by a party with probability of error $p\leq n^{-\frac{81}{16}}$. Theorem~\ref{th:rs} then ensures that $\Pi$ is successful except with probability $(32n^5n^{-\frac{81}{16}})^{\RC(\pi)}=n^{-\Omega(\RC(\pi))}$. In the following section, we shall show better designs of the broadcast algorithm BCast$()$, which take fewer than $O(\log n)$ rounds, while still maintaining a low enough $p$ to ensure that the simulation $\Pi$ fails with a vanishing probability of error only.

\section{Broadcast algorithms}\label{sec:bitex}
Theorem~\ref{th:rs} requires that any party receives a broadcast of a symbol from $S$ correctly, except with probability~$p$. Towards this goal we define \emph{bit-exchange} protocols that will be executed over each broadcast channel and allow the parties to reliably exchange their symbols consisting of $\log|S|$ bits with a probability high enough to allow the success of the simulation through Theorem~\ref{th:rs}.

\begin{definition}\label{def:bitExchange}
Let $n_1+1$ parties be connected by a broadcast channel, and assume each party $i\in[n_1+1]$ holds a bit $b_i\in\{0,1\}$.
A \emph{bit-exchange algorithm} with $R$ rounds and failure probability~$p$ is a protocol that performs $R$ fully-utilized broadcast rounds, after which all parties output $\{b_i\}_{i\in[n_1+1]}$, except with probability~$p$.\footnote{Notice that to use these bit exchange protocols as the algorithm BCast$()$ in Algorithms~\ref{alg:rsperip} and~\ref{alg:rscent}, we only need that any party is able to correctly decode the bit of any other party, except with probability $p$. Here, we are asking for something stronger, that all parties are able to decode all the bits correctly, except with probability $p$.}
\end{definition}

In this section, we introduce three different bit-exchange algorithms. When applied within the run of Algorithms~\ref{alg:rsperip} and~\ref{alg:rscent}, these three algorithms respectively give the complete simulation $\Pi$ which achieves the desired round complexity of the first three regimes in Theorem~\ref{th:main}. The fourth and last regime of Theorem~\ref{th:main} is dealt with in Section~\ref{sec:star}.

\subsection[The first regime]{The regime where $n_2=2^{O(\log n_1)}$}\label{sec:reg1}

In this regime, the bit exchange algorithm we shall use is the well-known Gallager scheme of \cite{Gal}, whose performance is characterized in the next theorem. 

\begin{theorem}[Section~III of \cite{Gal}]
\label{th:gal}
Let $n_1+1$ parties be connected over a single noisy broadcast channel with crossover probability~$\epsilon<1/2$. 
For any constant $c>0$ there exists a bit-exchange algorithm with   $O(\log \log n_1)$ rounds and failure probability~$n_1^{-c}$.  
\end{theorem}
We denote the scheme guaranteed by the above theorem by ExchBasic$()$. Note that the $O()$ notation hides the dependency on the constants~$\epsilon,c$.

We do not explicitly detail the pseudo-code of ExchBasic$()$ and refer the reader to~\cite{Gal} for the complete details. 
We note that the bit exchange algorithm of the second regime (see Algorithm~\ref{alg:ExchVar}) is a generalization of ExchBasic$()$. ExchBasic$()$ is obtained from Algorithm~\ref{alg:ExchVar} by simply changing the number of initial transmissions to $O(\log\log n_1)$ instead of $O(\log\log n_2)$, as well as changing the group size $m$ to $\Theta(\log n_1)$ instead of $\Theta(\log n_2)$. 

Now, observe that in the simulation $\Pi$ in Algorithms~\ref{alg:rsperip} and~\ref{alg:rscent}, the parties in $\bp_{i+0}$ need to exchange $\log|S|=O(1)$ bits each time BCast$()$ is called. We exchange each of these bits using $\log|S|$ runs of ExchBasic$()$, which costs us a total of $O(\log\log n_1)$ rounds. Thus, the simulation $\Pi$ in this regime, consists of $O(\RC(\pi)\cdot\log\log n_1)$ rounds. Taking the union bound over the $\log|S|=O(1)$ bits, Theorem~\ref{th:gal} ensures that for any constant $c>0$, after running ExchBasic$()$, any party in $\bp_{i+0}$ can recover all the $\log|S|$ bits broadcast by another party in $\bp_{i+0}$, except with probability $p=n_1^{-c}$. Plugging this value into Theorem~\ref{th:rs}, the simulation $\Pi$ will fail with a probability $(32n^5n_1^{-\frac{c}{16}})^{\RC(\pi)}$. Noting that in this regime $n_2=2^{O(\log n_1)}$, and hence, $n=n_1n_2+1\leq n_1^{c'}$, for some constant $c'\geq 0$. Plugging in this bound on $n$, the probability of failure of $\Pi$ is upper bounded by $(32n_1^{-\frac{c}{16}+5c'})^{\RC(\pi)}$. Choosing the constant $c>80c'$, thus yields a probability of error $n_1^{-\Omega(\RC(\pi))}\leq n^{-\Omega(\RC(\pi))}$, as $n\leq n_1^{c'}$. This completes the proof of the first regime in Theorem~\ref{th:main}.

\subsection[The second regime]{The regime where $n_2=2^{\Omega(\log n_1)}$ but $n_2=2^{O(n_1)}$}\label{sec:reg2}

We begin by specifying the exact boundaries for this regime. For this regime, we shall have $n_2$ lying in the range $2^{C_1\log n_1}\leq n_2\leq 2^{C_2n_1}$, where $C_1,C_2$ are specified later. In this regime, we will use a variant of the ExchBasic$()$ protocol as our bit exchange protocol, which we name ExchVar$()$ and depict in Algorithm~\ref{alg:ExchVar}.

The bit-exchange protocol works as follows. First partition the $n_1+1$ parties into groups of size $m<n_1+1$, and denote these groups as $A_l, l\in[L]$, where $L=\lceil\frac{n_1+1}{m}\rceil$.\footnote{Here, $\lceil\cdot\rceil$ denotes the ceiling function.} The partition can be arbitrary, but for the ease of exposition, we assume that $A_l$ consists of the parties $m(l-1)+1,\ldots,ml$.\footnote{Since $n_1+1$ may not be be a multiple of $m$, the final group $A_L$ may have less than $m$ parties. In such a scenario, we will simply assume that the final group has an additional $mL-n_1-1$ fictitious parties whose associated bit is zero.} In other words, the $j$th party of group $A_l$ holds the bit $b_{m(l-1)+j}$. For this bit exchange algorithm, we shall make use of the error correcting code $\ECC_\delta:\{0,1\}^m\to\{0,1\}^{mK}$, as given by Lemma~\ref{lem:KGDR}, with the choice $\delta<0.025$. The pseudocode for ExchVar$()$ is detailed in Algorithm~\ref{alg:ExchVar}.


\begin{algorithm}
\caption{ExchVar$()$}
\textbf{Input:} 
\begin{enumerate}[{1.}]
    \item Bits $\{b_i\}_{i\in[n_1+1]}$
    \item Constants $C>0, \rho>0$ to be specified later
\end{enumerate}
\begin{algorithmic}[1]
\Statex 
\Statex \textit{At the $j$th party of group $A_l$, $\forall j\in[m],l\in[L]$: }
\State Broadcast $b_{m(l-1)+j}$ for  $C\log\log n_2$ times \Comment{This costs $O(\log\log n_2)$ rounds}
\For{$j'\neq j$}
\State $s_{j,l}(j')\leftarrow$Broadcast bit from the $j'$th party in $A_l$ decoded by majority rule over $C\log\log n_2$ transmissions
\EndFor
\State $s_{j,l}(j)\leftarrow b_{m(l-1)+j}$ 
\State $t_{j,l}^{mK}\leftarrow\ECC_\delta(s_{j,l}^m)$
\State Broadcast $t_{j,l}((j-1)K+1))\circ\ldots\circ t_{j,l}(jK)$ with $\rho$ repetitions \Comment{This costs $O(1)$ rounds}
\For{$l'=1$ \textbf{to} $L$}
\For{$j'=1$ \textbf{to} $m$}
\State $u_{j,l|l'}\biggl((j'-1)K+1:j'K\biggr)\leftarrow$ The $K$ broadcast bits received $j'$th party in $A_{l'}$ decoded with majority rule over $\rho$ repetitions
\EndFor
\State $v_{j,l|l'}^m\leftarrow\ECC^{-1}(u_{j,l|l'}^{mK})$
\EndFor
\State \Return $v_{j,l}^{n_1+1}\triangleq(v_{j,l|l'}^m: l'\in[L])$
\end{algorithmic}
\label{alg:ExchVar}
\end{algorithm}

The following lemma bounds the probability of failure of ExchVar$()$.

\begin{lemma}\label{lem:ExchVar}
Let $n_1+1$ parties be connected via a noisy broadcast channel with crossover probability~$\epsilon<1/2$. 
Fix a constant $0<c<200$, and let $n_2$ be such that $n_2\geq 2^{C_1\log n_1}$ but $n_2\leq 2^{C_2n_1}$, for $C_1>2,C_2>0$ to be specified later. 
Then, there exists a bit-exchange algorithm with $O(\log\log n_2)$ rounds and failure probability at most $n_2^{-c}$.
\end{lemma}
\begin{IEEEproof}
We will show that ExchVar$()$, where each group $A_l$ consists of $m=\lceil C_3\log n_2\rceil$ parties, with $C_3=250(2c+1)$, is the required bit exchange algorithm. Choose $C_2=\frac{1}{100251}$. This ensures that as long as $c<200$, $C_2\leq \frac{1}{C_3+1}$, and hence, $n_1+1\geq m=\lceil C_3\log n_2\rceil$ since $n_2\leq 2^{n_1C_2}$. Thus, the above choice for $m$ is valid. For the ease of notation, in the following, we shall denote the $j$th party of group $A_l$ as $(j,l)$.

We prove Lemma~\ref{lem:ExchVar} in two steps. First, we show that the probability of the event that there exists some group $A_l$ such that more than $0.1m$ parties in $A_l$ have failed the decoding in the first stage, is upper bounded by $n_2^{-2c}$. To show this, define the random variable $X_{j,l}$ as follows:
\begin{equation*}
X_{j,l}=
\begin{cases}
1, & \text{if }s_{j,l}^m\text{ decoded by party }(j,l)\text{ in the first phase is correct}\\
0, & \text{otherwise}.
\end{cases}
\end{equation*}
By Lemma~\ref{lem:repeat}, the probability of $s_{j,l}(k)$ being a wrong decoding of the transmission by party $(k,l)$ in the first phase is upper bounded by $(2\sqrt{\epsilon(1-\epsilon)})^{C\log\log n_2}$. Choosing the constant $C$ suitably large (depending on $\epsilon$), this probability can be upper bounded by $(\log n_2)^{-2}$. Hence, by union bound over parties $(k,l), k\neq j$ in $A_l$, the probability $q_e$ that $X_{j,l}=0$ is upper bounded as $q_e\leq m(\log n_2)^{-2}\leq (\log n_2)^{-1}$, since $m=\Theta(\log n_2)$. Noting $\ex[X_{j,l}]=1-q_e$, we have
\begin{align*}
\pr\biggl(\sum_{j=1}^mX_{j,l}\leq 0.9m\biggr) & = \pr\biggl(\sum_{j=1}^mX_{j,l}\leq \ex[\sum_{j=1}^m X_{j,l}]\frac{0.9}{1-q_e}\biggr)\\
                               & \stackrel{(a)}{\leq} e^{-\frac{(0.1-q_e)^2m}{2(1-q_e)}} \\
                               & \stackrel{(b)}{\leq} e^{-0.025m}\\
                               & \stackrel{(c)}{\leq}n_2^{-(2c+1)},
\end{align*}
where $(a)$ follows from the Chernoff bound in Lemma~\ref{lem:chern}, $(b)$ follows by noting $\frac{(0.1-q_e)^2}{2(1-q_e)}\geq 0.025$, for all sufficiently low $q_e$ (i.e., for all sufficiently large $n_2$), and $(c)$ follows since $C_3=250(2c+1)$. Now, taking union bound over all groups $A_l,l\in[L]$, we have
\begin{equation}
\pr\biggl(\exists A_l: \sum_{j\in A_l}X_j\leq 0.9m\biggr)\leq Ln_2^{-(2c+1)}\leq n_2^{-2c}, \label{eq:groupsuccess}
\end{equation}
where the final inequality follows from the fact that $L\leq n_1\leq n_2$, since $n_2\geq 2^{C_1\log n_1}$, where $C_1>2$.

Next, we condition on the event that $\sum_{j=1}^l X_{j,l}>0.9m$, for all groups $A_l, l\in[L]$. Now, consider the party $(j,l)$ and any group $A_{l'}, l'\in[L]$. We begin by bounding the probability of error by party $(j,l)$ in decoding the bits $b_{m(l'-1)+1},\ldots,b_{ml'}$ possessed by the parties in $A_{l'}$. Note that there is always a set $J\subseteq [m]$ such that $|J|>0.9m$, and $X_{j',l'}=1$ for every $j'\in J$. In other words, for every $j'\in J$, the party $(j',l')$ has correctly decoded all the bits of the parties in $A_{l'}$ in the first phase. Now, define the random variable $Y_{j,l|j',l'}$ as follows:
\begin{equation*}
Y_{j,l|j',l'}=
\begin{cases}
1, & \text{if party }(j,l)\text{ obtains $u_{j,l|l'}((j'-1)K+1:j'K)$ after correct decoding}\\
0, & \text{otherwise}.
\end{cases}
\end{equation*}
We first argue that $\sum_{j'\in J}Y_{j,l|j',l'}> 0.8m$ implies that $(j,l)$ decodes all the bits possessed by the parties in $A_{l'}$ correctly, i.e., $v_{j,l|l'}^m=b_{m(l'-1)+1},\ldots,b_{ml'}$. To see this, note that by Lemma~\ref{lem:KGDR}, the distance of the error correcting code $\ECC_{\delta}$ is $(\frac{1}{2}-\delta)Km$. Thus, correct decoding is possible for the error correcting code $\ECC_{\delta}$ if $(j,l)$ receives at least $Km(1-(\frac{1}{4}-\frac{\delta}{2})-\frac{1}{2Km})$ bits broadcast by parties from $(j',l')$, for all $j'\in J$, correctly. Note that we only consider transmissions from $(j',l')$ with $j'\in J$, since only those are able to encode correctly as $X_{j',l'}=1$. Simple manipulations yield $Km(1-(\frac{1}{4}-\frac{\delta}{2})-\frac{1}{2Km})\leq Km(\frac{\delta}{2}+\frac{3}{4})$, and noting the choice $\delta<0.025$, we have $Km(\frac{\delta}{2}+\frac{3}{4})<0.8Km$. In other words, $(j,l)$ correctly decoding $u_{j,l|l'}((j'-1)K+1:j'K)$, for at least $0.8m$ parties $(j',l')$, $j'\in J$, ensures that $v_{j,l|l'}^m=b_{m(l'-1)+1},\ldots,b_{ml'}$. Hence, the event $\sum_{j'\in J}Y_{j,l|j',l'}\geq 0.8m$ implies that $v_{j,l|l'}^m=b_{m(l'-1)+1},\ldots,b_{ml'}$. We finish the proof by bounding the probability of the event $\sum_{j'\in J}Y_{j,l|j',l'}< 0.8m$.

Denote by $q_{e'}$ the probability that $(j,l)$ obtained $u_{j,l|l'}((j'-1)K+1:j'K)$ after incorrect decoding. By Lemma~\ref{lem:repeat}, we can ensure $q_{e'}$ to be arbitarily small, by fixing the number of repetitions $\rho$ large enough (depending on $K$ and $\epsilon$ only). Then $\ex[\sum_{j'\in J}Y_{j,l|j',l'}]=|J|(1-q_{e'})$. Therefore, by application of the Chernoff bound, we have
\begin{align*}
\pr\biggl(\sum_{j'\in J}Y_{j,l|j',l'}<0.8m\biggr) & =\pr\biggl(\sum_{j'\in J}Y_{j,l|j',l'}<\ex[\sum_{j'\in J}Y_{j,l|j',l'}]\frac{0.8m}{|J|(1-q_{e'})}\biggr)\\
                                           & \stackrel{(a)}{\leq}\pr\biggl(\sum_{j'\in J}Y_{j,l|j',l'}<\ex[\sum_{j'\in J}Y_{j,l|j',l'}]\frac{8}{9(1-q_{e'})}\biggr)\\
                                           & \stackrel{(b)}{\leq}e^{-\frac{0.9m(1-9q_{e'})^2}{162(1-q_{e'})}}\\
                                           & \stackrel{(c)}{\leq}e^{-\frac{m}{360}}\\
                                           & \stackrel{(d)}{\leq}n_2^{-(2c+1)},
\end{align*}
where $(a)$ follows from $|J|>0.9m$, $(b)$ uses the Chernoff bound in Lemma~\ref{lem:KGDR}, and $(c)$ uses the fact that $\frac{(1-9q_{e'})^2}{(1-q_{e'})}\geq 0.5$ for sufficiently small $q_{e'}$, and $(d)$ follows by noting $m=\lceil C_3\log n_2\rceil$ and $C_3=250(2c+1)$.

This ensures that party $(j,l)$ decodes all the bits possessed by the group $A_{l'}$ correctly except with probability $n_2^{-(2c+1)}$. Taking an union bound over all groups $A_{l'}, l'\in[L]$, and all parties $(j,l), j\in[m], l\in[L]$, we have that every party is able decode all the bits $b_1,b_2,\ldots,b_{n_1+1}$ correctly except with a probability $(n_1+1)Ln_2^{-(2c+1)}\leq n_1^{2.1}n_2^{-2c+1}\leq n_2^{-2c}$. The first inequality follows noting $L\leq n_1$, whereas the second inequality is obtained by choosing $C_1=2.1$ and noting $n_2\geq 2^{C_1\log n_1}$. 

Now note that this probability was obtained conditioned on the event that for all $A_l, l\in [L]$, we have $\sum_{j=1}^mX_{j,l}>0.9m$. Recall that this event does not occur with a probability $n_2^{-2c}$. Hence, by union bound, we have the probability of failure of ExchVar$()$ to be at most $2n_2^{-2c}\leq n_2^{-c}$, as required.
\end{IEEEproof}

The parties in $\bp_{i+0}$ run ExchVar$()$ for $\log|S|$ times to exchange each others tree coded symbols. Choosing $c=162$ in Lemma~\ref{lem:ExchVar}, each bit exchange has a failure probability of $n_2^{-162}$. Then, all of the $\log|S|$ bit exchanges are successful except with probability $|S|n_2^{-162}\leq n_2^{-161}$, as $|S|=O(1)$. In other words, we have $p=n_2^{-161}$. Next, observe that since $C_1>2$, and $n_2\geq 2^{C_1\log n_1}$, we have $n=n_1n_2+1\leq n_2^{2}$. Now, plugging in the value of $p$ in Theorem~\ref{th:rs}, the probability of failure of the simulation $\Pi$ is given by
\begin{gather*}
    (32n^5n_2^{-\frac{161}{16}})^{\RC(\pi)}\leq (32n_2^{10}n_2^{-\frac{161}{16}})^{\RC(\pi)}\leq n_2^{-\Omega(\RC(\pi)}\leq n^{-\Omega(\RC(\pi)},
\end{gather*}
where the first and final inequality follows using the fact that $n\leq n_2^2$. Finally, note that $\log|S|=O(1)$ executions of ExchVar$()$ costs $O(\log\log n_2)$ rounds, and hence the simulation $\Pi$ has a round complexity $\RC(\pi)O(\log\log n_2)$.

\subsection[The third regime]{The regime where $n_2=2^{\Omega(n_1)}$ but $n_2=2^{2^{O(\frac{n_1\log n_1}{\log\log n_1})}}$}\label{sec:reg3}

In this region, $n_2$ becomes too large (with respect to $n_1$), that exchanging bits via Gallager's method (ExchBasic$()$) does not yields a sufficient success probability. In order to reduce the failure probability to being $n_2^{-c}$ we will perform a large number of ExchBasic$()$ exchanges and take their majority. This ``repetition'' code will allow us  obtaining the required success probability by moderately increasing the overhead.
We name the bit exchange algorithm for this regime by ExchRepeat$()$ and depict it in Algorithm~\ref{alg:ExchRepeat}. 

\begin{algorithm}
\caption{ExchRepeat$()$}
\textbf{Input:} 
\begin{enumerate}[{1.}]
    \item Bits $\{b_q\}_{q\in[n_1+1]}$
    \item Constants $C>0$ to be specified later
\end{enumerate}
\begin{algorithmic}[1]
\State Run ExchBasic$()$ $C\frac{\log n_2}{\log n_1}$ times with the same input bits $\{b_q\}_{q\in[n_1+1]}$
\Statex
\Statex \textit{At $q$th party holding $b_q$, $\forall q\in[n_1+1]$:}
\Indent 
\For{$q'\neq q$}
\For{$r=1$ \textbf{to} $C\frac{\log n_2}{\log n_1}$}
\State $v_{q|q'}(r)\leftarrow$ Decoded bit (according to decoding rule of ExchBasic$()$) from party $q'$ for $r$th run of ExchBasic$()$
\EndFor
\State $u_{q|q'}\leftarrow$ The majority bit of $v_{q|q'}^{C\frac{\log n_2}{\log n_1}}$
\EndFor
\State \Return $(u_{q|q'}:q'\neq q)$
\EndIndent 
\end{algorithmic}
\label{alg:ExchRepeat}
\end{algorithm}



\begin{lemma}\label{lem:BE-R3}
Let $n_1+1$ parties be connected via a noisy broadcast channel with crossover probability~$\epsilon<1/2$. 
Fix a constant $c>0$ and let $n_2$ be such that $n_2=2^{\Omega(n_1)}$
Then, there exists a bit-exchange algorithm with $O(\frac{\log n_2}{\log n_1}\log\log n_1)$ rounds and failure probability at most $n_2^{-c}$.
\end{lemma}

\begin{IEEEproof} 
We begin by computing the probability of failure of ExchRepeat$()$. Firstly, fix $q,q'\in[n_1+1]$ with $q'\neq q$, and consider the probability that the $q$th party makes an error in the decoding $b_{q'}$ for the $r$th run of ExchBasic$()$. By Theorem~\ref{th:gal}, we can bound this probability by $n_1^{-1}$, at the cost of $O(\log\log n_1)$ rounds used in the $r$th run of ExchBasic$()$. In other words, for every $r\in[C\frac{\log n_2}{\log n_1}]$, $v_{q|q'}(r)$ can be viewed as an $n_1^{-1}$-noisy copy of $b_{q'}$. Hence, the probability that the $q$th party fails to decode $b_{q'}$, i.e., $u_{q|q'}\neq b_{q'}$, can be bounded using Lemma~\ref{lem:repeat} as follows.
\begin{align*}
    \text{Pr}(u_{q|q'}\neq b_{q'}) & \stackrel{(a)}{\leq} \biggl(2\sqrt{n_1^{-1}(1-n_1^{-1})}\biggr)^{C\frac{\log n_2}{\log n_1}}\\
                                   & \leq (2n_1^{-0.5})^{C\frac{\log n_2}{\log n_1}}\\
                                   & \stackrel{(b)}{\leq} n_1^{-0.4C\frac{\log n_2}{\log n_1}}\\
                                   & \leq n_2^{-0.4C},
\end{align*}
where $(a)$ is due to Lemma~\ref{lem:repeat} and $(b)$ follows by noting $2n_1^{-0.5}\leq n_1^{-0.4}$. Hence, taking a union bound over all $q$ and $q'$, we have that the probability of failure of ExchRepeat$()$ is bounded by
\begin{gather*}
    n_1(n_1+1)n_2^{-0.4C}\stackrel{(a)}{\leq}n_2^2n_2^{-0.4C}\stackrel{(b)}{\leq}n_2^{-c},
\end{gather*}
where $(a)$ follows by noting $n_2=2^{\Omega(n_1)}$, and $(b)$ follows by choosing the constant $C=\frac{c+2}{0.4}$. 

Finally, recall that each run of ExchBasic$()$ costs $O(\log\log n_1)$ rounds, and there are a total of $C\frac{\log n_2}{\log n_1}$ runs. Hence, ExchRepeat$()$ runs for a total of $O(\log\log n_1\frac{\log n_2}{\log n_1})$ rounds, as required. 
\end{IEEEproof} 

Similar to Sections~\ref{sec:reg1} and \ref{sec:reg2}, the parties in $\bp_{i+0}$ run ExchRepeat$()$ $\log|S|=O(1)$ times to exchange their tree code symbols. Noting that each run of ExchRepeat$()$ costs $O(\log\log n_1\frac{\log n_2}{\log n_1})$ rounds, we have that the round complexity of simulation $\Pi$ is $O(\RC(\pi)\frac{\log n_2\log\log n_1}{\log n_1})$. Also, choosing $c=162$ in Lemma~\ref{lem:BE-R3}, we have that a single run of ExchRepeat$()$ fails with a probability $n_2^{-162}$. Therefore, by union bound, $|S|=O(1)$ runs of ExchRepeat$()$ fails with probability at most $n_2^{-161}$, i.e., $p=n_2^{-161}$. Also, noting that $n_2=2^{\Omega(n_1)}$, we have $n=n_1n_2+1\leq n_2^2$. Plugging these in Theorem~\ref{th:rs}, we have that the simulation $\Pi$ succeeds except with a probability
\begin{gather*}
    (32n^5p^{\frac{1}{16}})^{\RC(\pi)}\leq (32n_2^{10-\frac{161}{16}})^{\RC(\pi)}\leq n_2^{-\Omega(\RC(\pi))}\leq n^{-\Omega(\RC(\pi))},
\end{gather*}
where the last inequality follows by recalling $n\leq n_2^2$. 

\section[Simultaneously simulating multiple rounds: the fourth regime]{Simultaneously simulating multiple rounds: the regime where $n_2=2^{2^{\Omega(\frac{n_1\log n_1}{\log\log n_1})}}$}\label{sec:star}

In the last regime, $n_2$ is extremely large with respect to~$n_1$. In the extreme case, where $n_1$ is constant or even equals~1, the problem degenerates into solving interactive coding over a star. Our simulation~$\Pi$ designated for this case follows technique from~\cite{BEGH18} for interactive coding over a star. 
The key idea is to simulate the noiseless protocol $\pi$ in large chunks of consecutive rounds. At the beginning of each chunk, the parties send~$p_0$ a list of all the possible messages they can send in~$\pi$ during the next chunk (as a function of the messages they will receive). $p_0$ collects all these lists and computes all the messages that $\pi$ would have sent over a noiseless network. Then, $p_0$ updates each party with the specific messages communicated to that party in the simulation of~$\pi$. We now describe this process (and its robust simulation) in greater detail.

We first illustrate an alternative way of viewing the underlying noiseless protocol $\pi$. For any party $p_{i,j}$, associate with it a $2^{n_1}$-ary tree $\tau_{i,j}$ of depth $\RC(\pi)$. Address the $2^{n_1}$ edges coming out of any node in the tree by $n_1$ bits. Any node $v\in\tau_{i,j}$ is identified by the ordered collection of the addresses of the edges in the unique path from the root to~$v$. 
Also, each node~$v$ of the tree is associated with a bit~$b_v$ that is determined by the protocol~$\pi$ and the input $x_{i,j}$ to $p_{i,j}$ as follows. Consider any node~$v$ in level $r-1$ whose address is $m_{i,j}^{r-1}\in\{0,1\}^{n_1(r-1)}$. Set $b_v=\pi(x_{i,j},m_{i,j}^{r-1})$. One can think of $b_v$ as the bit that $p_{i,j}$ sends in $\pi$ at the $r$-th round, given that its input is~$x_{i,j}$ and that during the $(r-1)$-th first rounds of~$\pi$ it has received the messages $m_{i,j}^{r-1}$ from~$\bp_{i+0}$.
Similarly, for the central party~$p_0$ we define a $2^{n_1n_2}$-ary tree $\tau_0$ of depth $\RC(\pi)$, and associate with the node~$v$ in $\tau_0$ addressed by $m_0^{r-1}$, a string~$b_v$ of $n_2$~bits, defined by $b_v=\pi(x_0,m_0^{r-1})$. 
As in the case of~$p_{i,j}$, the string $b_v$ can be thought of as all the bits $p_0$ broadcasts in~$\pi$ to all $\bp_i$, $i\in[n_2]$, given its input is~$x_0$ and that its received communication up to round~$r-1$ is~$m_0^{r-1}$.

From the previous discussion, it is thus easy to see that a run of $\pi$ corresponds to the parties finding a path down their associated tree. For party $p_q$, its received transcript $\trec_q$ thus corresponds to a leaf in $\tau_q$. Also, the sent transcript of party $p_q$, $\tsent_q$, consisting of the bits it sent during the run of $\pi$, is simply the bits $b_v$ associated with the nodes in $\tau_q$ along the path from the root to $\trec_q$. 

Next, we describe $\tilde{\pi}$, the $k$-chunk version of the original noiseless protocol $\pi$. Similar to~$\pi$, in absence of noise, after running $\tilde{\pi}$, all parties $p_q$ are able to output their transcripts $\tsent_q$, $\trec_q$. However, instead of advancing level-by-level in $\tau_0,\tau_{i,j}$ as in~$\pi$, the parties now advance in chunks of $k$ consecutive levels. To simulate a chunk of size $k$, first the parties in $\cup_{i\in[n_2]}\bp_i$, broadcasts the bits associated with all the nodes in the next $k$ levels of their respective trees, in a predetermined order. In other words, all non-central parties describe all of their possible broadcasts for the next $k$ levels. On receiving these, the central node $p_0$ has the relevant information to extend its path in $\tau_0$ by $k$ levels. Next, it broadcasts the bits associated with these newly added $k$ nodes to their respective broadcast links. Following these broadcasts, any party $p_{i,j}$ sees all the possible broadcasts by all the other parties in $\bp_i$, as well as the exact messages from $p_0$ for the next $k$ levels. In other words, $p_{i,j}$ now has enough information to extend its path in $\tau_{i,j}$ by $k$ levels. This process is repeated until all levels are exhausted. A more formal description follows. 

The scheme~$\tilde{\pi}$ consists of $\lceil \frac{1}{k}\RC(\pi)\rceil$ chunks, with each chunk simulating $k$ levels.\footnote{Note that $\RC(\pi)$ need not be a multiple of $k$, and hence $k\lceil\frac{1}{k}\RC(\pi)\rceil$ may exceed $\RC(\pi)$. To account for these additional levels, we zero-pad the original protocol $\pi$.} Formally, assume after running $l$ chunks, party $p_{i,j}$ has received messages $(\tilde{m}_{i,j'}^l: j'\neq j)$ sent by the other parties in $\bp_i$, and $\tilde{m}_{0|i}^l$ sent by $p_0$. Similarly, the central party has received messages $(\tilde{m}_{i,j}^l:i\in[n_2],j\in[n_1])$ from all other parties. Define functions $\cpath_{i,j}$ and $\cpath_0$, which compute the correct path down the trees $\tau_{i,j}$ and $\tau_0$ based on the received messages in a chunk-by-chunk basis, as described in the previous paragraph. Hence, we have $\cpath_{i,j}\biggl(\tilde{m}_{0|i}^l, \tilde{m}_{i,j'}^l: j'\neq j\biggr)=m_{i,j}^{kl}$ and $\cpath_0\biggl(\tilde{m}_{i,j}^{l+1}:i\in[n_2],j\in[n_1]\biggr)=m_0^{kl}$, where recall that $m_{i,j}^{kl}$ (resp., $m_0^{kl}$) is the received transcript of party $p_{i,j}$ (resp., $p_0$) after $kl$ rounds of $\pi$. The party $p_{i,j}$ then broadcasts $\tilde{m}_{i,j}(l+1)\triangleq\NxtLvls_{i,j}(x_{i,j},m_{i,j}^{kl})$, which consists of all the bits $b_v$ associated with nodes in the first $k$ levels of the subtree of $\tau_{i,j}$ rooted at $m_{i,j}^{kl}$. Note that since $\tau_{i,j}$ is a $2^{n_1}$-ary tree, $\tilde{m}_{i,j}(l+1)$ consists of $\Theta(2^{n_1k})$ bits. During these rounds, the central party simply sends zeros. On receiving these, the central party is able to extend its path by next $k$ levels to $m_0^{k(l+1)}$. In other words, we have $\cpath_0\biggl(\tilde{m}_{i,j}^{l+1}:i\in[n_2],j\in[n_1]\biggr)=m_0^{k(l+1)}$. The party $p_0$ then obtains $\tilde{m}_0(l+1)=\tilde{m}_{0|1}(l+1)\ldots\tilde{m}_{0|n_2}(l+1)\triangleq\NxtLvls_0(x_0,m_0^{k(l+1)})$,\footnote{Note that the argument in $\NxtLvls_0$ is $m_0^{k(l+1)}$, as opposed to $m_{i,j}^{(kl)}$ in $\NxtLvls_{i,j}$. This is because $p_0$ broadcasts after receiving messages from all other parties, and hence after extending its path by $k$ levels.} where $\tilde{m}_{0|i}(l+1)$ is the collection of the $i$th bits of the bit strings associated with the nodes along the path from $m_0^{kl}$ to $m_0^{k(l+1)-1}$ in $\tau_0$. In other words $\tilde{m}_{0|i}(l+1)$ consists of $k$ bits, which $p_0$ broadcasts to $\bp_i$. During these rounds, the parties $p_{i,j}$ broadcasts zeros. On receiving $\tilde{m}_{0|i}(l+1)$, the parties $p_{i,j}$ are able to compute $m_{i,j}^{k(l+1)}=\cpath_{i,j}\biggl(\tilde{m}_{0|i}^{l+1}, \tilde{m}_{i,j'}^{l+1}: j'\neq j\biggr)=m_{i,j}^{k(l+1)}$. The pseudocode of the $k$-chunked scheme $\tilde{\pi}$ is available in Algorithm~\ref{alg:chunk}. 

\begin{algorithm}[ht]
\caption{$k$-chunk scheme $\tilde{\pi}$}

\begin{algorithmic}[1]
\Statex \textbf{Input:} $k>1$, $\pi$, $x=(x_0,x_{i,j}: i\in[n_2], j\in[n_1])$ ($x$ is distributed among parties)
\Statex
\State \textbf{Initialize} 
Let $\tau_{i,j}$ (resp., $\tau_0$) be the tree-description of $\pi$ for party $p_{i,j}$ ($p_0$) assuming the input~$x_{i,j}$ ($x_0$). 
For any $i\in[n_2], j\in[n_1]$ set $m_{i,j}^0=$ root of $\tau_{i,j}$; $m_0^0=$ root of $\tau_0$.
\Statex
\For{$l=1$ \textbf{to} $\lceil\frac{1}{k}\RC(\pi)\rceil$}
\State \textbf{First part}\strut
\State \textit{At parties $p_{i,j}, \forall i\in[n_2], j\in[n_1]$, \textbf{do}:}
\State \hspace{\algorithmicindent} 
$\tilde{m}_{i,j}(l) \gets\NxtLvls_{i,j}(x_{i,j},m_{i,j}^{k(l-1)}$  
\State \hspace{\algorithmicindent} Broadcast $\tilde{m}_{i,j}(l)$ to $\bp_{i+0}$
\State \textit{At party $p_0$, \textbf{do}:}
\State \hspace{\algorithmicindent} Broadcast `0'.
\Statex

\State \textbf{Second part} \Comment{This occurs after all parties have executed the first part}
\State \textit{At party $p_0$, \textbf{do}:}
\State \hspace{\algorithmicindent}
$m_0^{kl}\leftarrow\cpath_0\biggl(\tilde{m}_{i,j}^l:i\in[n_2],j\in[n_1])\biggr)$
\State \hspace{\algorithmicindent} $\tilde{m}_0(l)\gets\NxtLvls_0(x_0,m_0^{kl})$ 
\State \hspace{\algorithmicindent} Broadcast $\tilde{m}_{0|i}(l)$ to $\bp_i, \forall i\in[n_2]$
\State \textit{At parties $p_{i,j}, \forall i\in[n_2], j\in[n_1]$, \textbf{do}:}
\State \hspace{\algorithmicindent}
Broadcast `0'.
\State \hspace{\algorithmicindent} (at the end of this part:) 
\Statex
\hspace{\algorithmicindent}\hspace{\algorithmicindent}$m_{i,j}^{kl}\gets\cpath_{i,j}\biggl(\tilde{m}_{0|i}^{l},\tilde{m}_{i,j'}^{l}: j'\neq j\biggr)$
\EndFor
\Statex
\State $\trec_0\leftarrow m_0^{\RC(\pi)}, \trec_{i,j}\leftarrow m_{i,j}^{\RC(\pi)}, \forall i\in[n_2], j\in[n_1]$
\State $\tsent_0\leftarrow$ Bits associated with nodes along the path to $m_0^{\RC(\pi)}$ in $\tau_0$ \strut
\State $\tsent_{i,j}\leftarrow$ Bits associated with nodes along the path to $m_{i,j}^{\RC(\pi)}$ in $\tau_{i,j}$, $\forall i\in[n_2], j\in[n_1]$
\State Parties $p_{i,j}, \forall i\in[n_2], j\in[n_1]$ returns their own transcripts $\trec_{i,j},\tsent_{i,j}$
\Statex Party $p_0$ returns its own transcripts $\trec_0,\tsent_0$
\end{algorithmic}
\label{alg:chunk}
\end{algorithm}

The simulation $\Pi$ for this regime will be simulating $\tilde{\pi}$ instead of $\pi$, with $k=\lceil\frac{\log\log n_2}{n_1}\rceil$. Notice that this choice for $k$ is valid in the regime $n_2=2^{2^{\Omega(\frac{n_1\log n_1}{\log\log n_1})}}$, as $\log\log n_2>n_1$. The round complexity and the failure probability of $\Pi$ is stated in the following theorem.
\begin{theorem}
\label{th:star}
Let $n_2=2^{2^{\Omega(\frac{n_1\log n_1}{\log\log n_1})}}$ and fix $\epsilon>0$. Consider the network of intersecting broadcast links with $n=n_1n_2+1$ parties affected by $\epsilon$-bit flip noise. Then, for every noiseless protocol $\pi$ with $\RC(\pi)=R$, there is a simulation $\Pi$, which runs for $O(R\cdot\frac{n_1\log n_2}{\log\log n_2})$ rounds, and fails with a vanishing probability.
\end{theorem}

We conclude this section by describing the simulation $\Pi$, which involves the RS simulation of $\tilde{\pi}$. The proof of Theorem~\ref{th:star} is therefore similar to that of Theorem~\ref{th:rs}, but involves a few changes arising due to a couple of reasons. Firstly, in $\tilde{\pi}$ the number of bits sent by $p_0$ in a step differs from that of the other parties. Secondly, a step in $\tilde{\pi}$ involves first the parties in $\bigcup_{i\in[n_2]}\bp_i$ making their transmissions, and based on that $p_0$ transmits. This is different from $\pi$, where, within each step, all transmissions are simultaneous. The complete proof of Theorem~\ref{th:star} appears in Appendix~\ref{app:star} in the supplementary material. 

\begin{algorithm}
\caption{The simulation $\Pi$ for the regime $n_2=2^{2^{\Omega(\frac{n_1\log n_1}{\log\log n_1})}}$}
\label{alg:star}
\textbf{input:} 
\begin{enumerate}[{1.}]
\item $k=\lceil\frac{\log\log n_2}{n_1}\rceil$
\item Protocol $\tilde{\pi}$ with zero-padding after $\RC(\tilde{\pi})$ transmissions
\item Input $x=(x_0,x_{i,j}: i\in[n_2],j\in[n_1])$
\end{enumerate}
\begin{algorithmic}[1]
\State \textbf{initialize} $w_{i,j}\leftarrow\emptyset, s_{i,j}\leftarrow\emptyset,\;\; \forall i\in[n_2], j\in[n_1]$, $w_0\leftarrow\emptyset, s_0\leftarrow\emptyset$
\Statex
\For{$r=1$ \textbf{to} $2\lceil\frac{1}{k}\RC(\pi)\rceil$}
\State \textbf{First part}
\State \textit{At party $p_{i,j}, \forall i\in[n_2],j\in[n_1]$}: Decode $\hat{w}_{i,j}^{r-1}$ 
\Indent 
\State $u_{i,j}\leftarrow$ ConsCheckChunked$(\tilde{\pi},(i,j),x_{i,j},\hat{w}_{i,j}^{r-1},w_{i,j}^{r-1},s_{i,j}^{r-2})$ 
\If{$u_{i,j}=1$} \Comment{Consistency checks}                         
\State $s_{i,j|j'}(r-1)\leftarrow\hat{w}_{i,j|k}(r-1), \; \; \;  \forall 0\leq j'\leq [n_1], j'\neq j$
\State $w_{i,j}(r)\leftarrow\NxtLvls_{i,j}\biggl(x_{i,j},\cpath_{i,j}(\sigma_{i,j}^{(r)})\biggr)$
\Else
\State $s_{i,j}(r-1)\leftarrow\BK^{n_1}$, $w_{i,j}(r)\leftarrow\BK$      \Comment{Moving back by one step, if inconsistent}
\EndIf
\State Broadcast $\ECC_\nc(\cT_\nc(w_{i,j}^r)(r))$ to $\bp_{i+0}$       \Comment{$\ECC_\nc$ is applied on the $r$th symbol of $\cT_\nc(w_{i,j}^r)$}
\EndIndent
\State \textit{At party $p_0$:}
\Indent 
\State Broadcast `0'
\EndIndent
\Statex
\State \textbf{Second part} \Comment{This occurs after concluding the first part at all parties}
\State \textit{At party $p_0$}: Decode $\hat{w}_0^{r}$ 
\Indent 
\State $u_0\leftarrow$ ConsCheckChunked$(\tilde{\pi},0,x_0,\hat{w}_0^{r},w_0^{r-1},s_0^{r-1})$ 
\If{$u_0=1$} \Comment{Consistency checks}                         
\State $s_{0|i,j}(r)\leftarrow\hat{w}_{0|i,j}(r), \; \; \;  \forall i\in[n_2], j\in[n_1]$
\State $w_0(r)\leftarrow\NxtLvls_0\biggl(x_0,\cpath_0(\sigma_0^{(r)})\biggr)$
\Else
\State $s_0(r)\leftarrow\BK^{n_1n_2}$, $w_0(r)\leftarrow\BK^{n_2}$      \Comment{Moving back by one step, if inconsistent}
\EndIf
\State Broadcast $\ECC_\cc(\cT_\cc(w_{0|i}^r)(r))$ to $\bp_i$, $\forall i\in[n_2]$        \Comment{$\ECC_\cc$ is applied on the $r$th symbol of $\cT_\cc(w_{0|i}^r)$}
\EndIndent
\State \textit{At parties $p_{i,j}, \forall i\in[n_2],j\in[n_1]$:}
\Indent 
\State Broadcast `0'
\EndIndent
\EndFor

\Statex

\State \textit{At party $p_{i,j}, \forall i\in[n_2],j\in[n_1]$}: 
\Indent 
\State Decode $\hat{w}_{i,j}^{2\lceil\frac{1}{k}\RC(\pi)\rceil}$ \Comment{Decoding the final transmission from the other parties at $p_{i,j}$}
\State $u_{i,j}\leftarrow$ ConsCheckChunked$(\tilde{\pi},(i,j),x_{i,j},\hat{w}_{i,j}^{2\lceil\frac{1}{k}\RC(\pi)\rceil},w_{i,j}^{2\lceil\frac{1}{k}\RC(\pi)\rceil},s_{i,j}^{2\lceil\frac{1}{k}\RC(\pi)\rceil-1})$
\If{$u_{i,j}=1$}
\State $s_{i,j|j'}(2\lceil\frac{1}{k}\RC(\pi)\rceil)\leftarrow\hat{w}_{i,j|j'}(2\lceil\frac{1}{k}\RC(\pi)\rceil), \;   \forall 0\leq j'\leq [n_1], j'\neq j$
\Else
\State $s_{i,j}(2\lceil\frac{1}{k}\RC(\pi)\rceil)\leftarrow\BK^{n_1}$
\EndIf
\State $\hat{\text{T}}^{(\text{s})}_{i,j}\leftarrow\parse(w_{i,j}^{2\lceil\frac{1}{k}\RC(\pi)\rceil})$, $\hat{\text{T}}^{(\text{r})}_{i,j}\leftarrow\cpath_{i,j}\biggl(\parse(s_{i,j|j'}^{2\lceil\frac{1}{k}\RC(\pi)\rceil}):0\leq j'\leq n_1, j'\neq j\biggr)$\Comment{Calculate estimated transcript}
\State\Return $\trec_{i,j},\tsent_{i,j}$
\EndIndent
\Statex\Statex
\State \textit{At party $p_0$}: 
\Indent 
\State $\hat{\text{T}}^{(\text{s})}_0\leftarrow\biggl(\parse(w_{0|i}^{2\lceil\frac{1}{k}\RC(\pi)\rceil}):i\in[n_2]\biggr)$, $\hat{\text{T}}^{(\text{r})}_0\leftarrow\cpath_0\biggl(\parse(s_{0|i,j}^{2\lceil\frac{1}{k}\RC(\pi)\rceil}):i\in[n_2], j\in[n_1]\biggr)$ \Comment{Calculate estimated transcript}
\State\Return $\trec_0,\tsent_0$
\EndIndent
\end{algorithmic}
\end{algorithm}

\begin{algorithm}
\caption{ConsCheckChunked() for step $r$}
\label{alg:consstar}
\begin{algorithmic}[1]
\State \textit{At party $p_{i,j}, \forall i\in[n_2],j\in[n_1]$:}
\Indent 
\State $u\leftarrow 1$
\State $\hat{z}_{i,j|j'}^{(r)}\leftarrow \parse(\hat{w}_{i,j|j'}^{r-1}), \forall 0\leq j'\leq [n_1],j'\neq j$
\State $\hat{\ell}_{i,j}^{(r)}\leftarrow\min_{\substack{0\leq j'\leq n_1\\ j'\neq j}}|\hat{z}_{i,j|j'}^{(r)}|$
\If{$\hat{\ell}_{i,j}^{(r)}\leq \ell_{i,j}^{(r-1)}$}
\State $u\leftarrow 0$
\Else
\For{$l=1$ \textbf{to} $\ell_{i,j}^{(r-1)}$}
\Comment{Checking if estimated transcripts agree with $\pi$}
\State $u\leftarrow u\wedge\biggl(\wedge_{\substack{0\leq j'\leq n_1\\ j'\neq j}}\mathbf{1}\biggl\{\hat{z}_{i,j|j'}^{(r)}(l)=\sigma_{i,j|j'}^{(r-1)}(l)\biggr\}\biggr)\wedge\mathbf{1}\{z_{i,j}^{(r-1)}(l+1)=\NxtLvls_{i,j}(x_{i,j},\cpath_{i,j}(\sigma_{i,j}^{(r-1)l}))\}$ 
\EndFor
\EndIf
\State \Return $u$
\EndIndent 
\Statex
\State \textit{At party $p_0$:}
\Indent 
\State $u\leftarrow 1$
\State $\hat{z}_{0|i,j}^{(r)}\leftarrow \parse(\hat{w}_{0|i,j}^{r}), \forall i\in[n_2], j\in[n_1]$
\State $\hat{\ell}_0^{(r)}\leftarrow\min_{i\in[n_2], j\in[n_1]}|\hat{z}_{0|i,j}^{(r)}|$
\If{$\hat{\ell}_0^{(r)}\leq \ell_0^{(r-1)}$}
\State $u\leftarrow 0$
\Else
\For{$l=1$ \textbf{to} $\ell_0^{(r-1)}$}
\Comment{Checking if estimated transcripts agree with $\pi$}
\State $u\leftarrow u\wedge\biggl(\wedge_{i\in[n_2], j\in[n_1]}\mathbf{1}\biggl\{\hat{z}_{0|i,j}^{(r)}(l)=\sigma_{0|i,j}^{(r-1)}(l)\biggr\}\biggr)\wedge\mathbf{1}\{z_0^{(r-1)}(l)=\NxtLvls_0(x_0,\cpath_0(\sigma_0^{(r-1)l}))\}$ 
\EndFor
\EndIf
\State \Return $u$
\EndIndent

\end{algorithmic}
\end{algorithm}

In what follows, we take $k=\lceil\frac{\log\log n_2}{n_1}\rceil$. As in the RS scheme, the parties run the protocol on an alphabet consisting of the $\BK$ symbol added to the original alphabet in the noiseless case. Since $p_0$ uses a different alphabet in $\tilde{\pi}$ than the remaining parties, it will also use a different alphabet in the simulation. More precisely, the parties $p_{i,j}$ communicate using the alphabet $\Sigma_\nc=\{0,1\}^{\Theta(2^{n_1k})}\cup\{\BK\}$, while the party $p_0$ uses the alphabet $\Sigma_\cc=\{0,1\}^k\cup\{\BK\}$. In order to transmit information, the parties $p_{i,j}$ (resp., $p_0$) use a tree code $\cT_\nc:\Sigma_\nc^*\to S_\nc^*$ (resp., $\cT_\cc:\Sigma_\cc^*\to S_\cc^*$), of distance $\alpha=0.5$. By Lemma~\ref{lem:ex}, we have that $S_\nc$ (resp., $S_\cc$) consists of $\Theta(2^{n_1k})$ (resp., $\Theta(k)$) bits. Minimum distance decoding, i.e., the decoding rule given in \eqref{eq:dec}, is used to decode both $\cT_\nc$ and $\cT_\cc$. The parties $p_{i,j}$ (resp., $p_0$) further employ an error correcting code $\ECC_\nc:S_\nc\to\{0,1\}^{O(\log n_2)}$ (resp., $\ECC_\cc:S_\cc\to\{0,1\}^{O(\log n_2)}$) to encode the tree code symbols into strings of bits before broadcasting them. The following lemma, a combination of Lemmas~\ref{lem:ecc} and~\ref{lem:repeat}, bounds the probability of decoding error of $\ECC_\nc$ and $\ECC_\cc$.
\begin{lemma}\label{lem:eccex}
Fix any $\epsilon>0$, and let $n_2=2^{2^{\Omega(\frac{n_1\log n_1}{\log\log n_1})}}$ and $k=\lceil\frac{\log\log n_2}{n_1}\rceil$. Let $m_\nc=\Theta(2^{n_1k})=\Theta(\log n_2)$ and $m_\cc=\Theta(k)$. Then, there exists error correcting codes $\ECC_\nc:\{0,1\}^{m_\nc}\to\{0,1\}^{O(\log n_2)}$ and $\ECC_\cc:\{0,1\}^{m_{\cc}}\to\{0,1\}^{O(\log n_2)}$, which can be correctly decoded in presence of $\epsilon$-bit flip noise, except with a probability $n_2^{-52}$. 
\end{lemma}
\begin{IEEEproof}
We show the case for $\ECC_\nc$. The proof for $\ECC_\cc$ is exactly similar by noting $k<\log n_2$. Firstly, by Lemma~\ref{lem:ecc}, there exists an error correcting code $\ECC:\{0,1\}^{m_\nc}\to\{0,1\}^{\Theta(m_\nc)}$, with probability of error $2^{-cm_\nc}$, for some constant $c>0$. Noting that $m_\nc=\Theta(\log n_2)$, we see that the output length of this code matches the requirement of $\ECC_\nc$. Furthermore, $m_\nc\leq c'\log n_2$, for some $c'>0$, since $m_\nc=\Theta(\log n_2)$. Thus, if $cc'\geq52$, the probability of decoding error of $\ECC$ is $2^{-cm_\nc}\leq n_2^{-cc'}\leq n_2^{-52}$, and we can simply use this $\ECC$ as $\ECC_\nc$. 

If however, $cc'<52$, then we concatenate some constant number $\rho>0$ copies of $\ECC$ to get $\ECC_\nc$. The output length then still remains $O(\log n_2)$. The decoding rule is as follows. We first decode all the $\rho$ copies of $\ECC$ using its decoding rule, to get $\rho$ estimated messages of $m_\nc$ bits. Denote these messages by $y_i\in\{0,1\}^{m_\nc}, i\in[\rho]$. Then, for any $j\in[m_\nc]$, declare the $j$th bit of the decoded message to be the majority bit among $(y_i(j))_{i\in[\rho]}$. Next, we analyze the probability of error. Consider the $j$th bit for some $j\in[m_\nc]$. Firstly, by Lemma~\ref{lem:ecc}, we have that $y_i(j)$ is an $n_2^{-cc'}$-noisy copy of the true $j$th bit. Then, by Lemma~\ref{lem:repeat}, we can bound the probability of error in decoding the $j$th bit by $(2\sqrt{n_2^{-cc'}(1-n_2^{-cc'})})^\rho\leq n_2^{-0.4cc'\rho}$. Choose $\rho>\frac{53}{cc'}$, to ensure that the $j$th bit is decoded correctly except with probability $n_2^{-53}$. Therefore, taking union bound over all $m_\nc\leq c'\log n_2$ bits, the probability of decoding error of $\ECC_\nc$ is upper bounded by $(c'\log n_2) n_2^{-53}\leq n_2^{-52}$.   
\end{IEEEproof}
We shall extend the definition of the parse operation in Definition~\ref{def:parse} to include larger alphabets. The operation still remains the same, i.e., deleting every $\BK$ symbol and the first non-$\BK$ symbol preceding it. 

As in the RS scheme described in Section~\ref{sec:RS}, we shall continue to use the same symbols $w_{i,j}^{r-1}\in\Sigma_\nc^{r-1}, w_0^{r-1}\in\Sigma_\cc^{n_2\cdot(r-1)}, s_{i,j}^{r-2}\in\Sigma_\cc^{r-2}\times\Sigma_\nc^{(n_1-1)\cdot(r-2)}, s_0^{r-1}\in\Sigma_\nc^{n_1n_2\cdot(r-1)}$ to denote the transmitted and the estimated strings at the various parties after the $(r-1)$th step.\footnote{Notice that after $(r-1)$th step $p_0$ has $r-1$ symbols, instead of $r-2$ symbols. This is because in the simulation $\Pi$, the parties $p_{i,j}$ transmit relevant information at the beginning of a step, which are first decoded by $p_0$ before sending its relevant transmission. Thus, in each step $p_0$ has one more decoded symbol than $p_{i,j}$.} Similarly, we retain the same notations of the parsed versions of these strings, i.e., $z_{i,j}^{(r-1)}=\parse(w_{i,j}^{r-1})$, $z_{0|i}^{(r-1)}=\parse(w_{0|i}^{r-1})$, $\sigma_{i,j|j'}^{(r-1)}=\parse(s_{i,j|j'}^{r-2})$, and $\sigma_{0|i,j}^{(r-1)}=\parse(s_{0|i,j}^{r-1})$, which serve as the estimated sent and received transcripts for the respective parties.\footnote{Again notice the extra symbol decoded by $p_0$.} Also, recall the notations, $\ell_{i,j}^{(r-1)}=|z_{i,j}^{(r-1)}|-1=|\sigma_{i,j|j'}^{(r-1)}|$ for all $j'\neq j$, and $\ell_0^{(r-1)}=|z_{0|i}^{(r-1)}|=|\sigma_{0|i,j}^{(r-1)})|$, for all $i\in[n_2], j\in[n_1]$. Similar to Algorithms~\ref{alg:rsperip} and~\ref{alg:rscent}, the simulation $\Pi$ (see Algorithm~\ref{alg:star}) also ensures that $\sigma_{i,j|j'}^{(r-1)}$, for all $j'\neq j$, are of the same length. Moreover, the length is one less than the length of $z_{i,j}^{(r-1)}$. This follows from the fact that the simulation $\Pi$ always appends a $\BK$ symbol to $w_{i,j}$ and $s_{i,j|j'}, j'\neq j$ at exactly the same steps. For party $p_0$, there is a slight change. Notice that now $|z_{0|i}^{(r-1)}|=|\sigma_{0|i,j}^{(r-1)}|$, instead of $|z_{0|i}^{(r-1)}|=|\sigma_{0|i,j}^{(r-1)}|+1$. This is again due to the fact that $p_0$ decodes in the middle of a step now, and thus sees one extra symbol than $p_{i,j}$.

The simulation $\Pi$ is detailed in Algorithm~\ref{alg:star}. It consists of $2\lceil\frac{1}{k}\RC(\pi)\rceil$ steps, with each step being divided into two parts as in $\tilde{\pi}$. At the beginning of the $r$th step, the parties $p_{i,j}$ collects the $r-1$ symbols received so far from the other parties in $\bp_{i+0}$, and decodes them first using the decoding rules for $\ECC_\cc$ (for the symbols coming from $p_0$) and $\ECC_\nc$ (for symbols coming from $p_{i,j'}, j'\neq j$), followed by the minimum distance decoding for the respective tree codes, to obtain the decoded string $\hat{w}_{i,j}^{r-1}=(\hat{w}_{i,j|j'}^{r-1}:0\leq j'\leq n_1, j'\neq j)\in\Sigma_\cc^{r-1}\times\Sigma_\nc^{(n_1-1)\cdot(r-1)}$. The party then checks if the decoded string is consistent with its estimated transcript. The consistency check procedure is similar to Algorithm~\ref{alg:cons}, but now instead of checking the transcript level by level, we check it chunk by chunk. The consistency check procedure is detailed in Algorithm~\ref{alg:consstar}. If the decoded string passes the consistency check, the party $p_{i,j}$ uses the function $\NxtLvls_{i,j}$ to send the relevant information from the next chunk.\footnote{We shall zero-pad $\tilde{\pi}$. Thus, if there are no further symbols to be sent, then send zeros.} Else, it decides to send the back symbol $\BK$. Once decided, the symbol is first encoded via the tree code $\cT_\nc$, followed by the error correcting code $\ECC_\nc$, and then broadcast to $\bp_{i+0}$. During these rounds party $p_0$ simply broadcasts zeros. On receiving these broadcasts, $p_0$ collects all the $r$ symbols it has received so far from all the other parties, and decodes the strings received from each party to obtain $\hat{w}_0^r=(\hat{w}_{0|i,j}^r:i\in[n_2],j\in[n_1])$. The party $p_0$ now checks the consistency of the decoded strings with its estimated transcripts according to Algorithm~\ref{alg:consstar}. If the decoded strings are consistent with its observation so far according to $\tilde{\pi}$, the party $p_0$ advances its path by one chunk, and decides the next set of bits to be broadcast to each $\bp_i$ using $\NxtLvls_0$. Else, if the decoded strings are inconsistent, $p_0$ decides to send the $\BK$ symbol to every $\bp_i$. The symbols to be broadcast are then encoded by the tree code $\cT_\cc$ and the error correcting code $\ECC_\cc$, and then sent to the respective links. During these rounds, the parties other than $p_0$ simply broadcasts zeros. 

On completing all the steps, the parties output restrictions of their estimated transcripts to the first $\RC(\pi)$ locations. Note that in each step, the broadcasts consist of $O(\log n_2)+O(\log n_2)=O(\log n_2)$ rounds. Thus, the round complexity of $\Pi$ is $2\lceil\frac{1}{k}\RC(\pi)\rceil O(\log n_2)=\RC(\pi)O(\frac{n_1\log n_2}{\log\log n_2})$, since $k=\lceil\frac{\log\log n_2}{n_1}\rceil$. To prove Theorem~\ref{th:star}, it remains to show that the simulation $\Pi$ has a vanishing probability of error, which is detailed in Appendix~\ref{app:star} in the supplementary material.

\bibliographystyle{IEEEtran}
\bibliography{CliquesStar}

\newpage

\begin{appendices}

\section{Proof of Theorem~\ref{th:rs}}\label{app:rs}

Consider any party $p_q$ after step $r$. We shall define the real progress at $p_q$ after step $r$ as the length upto which the estimated transcript of $p_q$ agrees with the execution of $\pi$. More formally, define
\begin{equation}
\rp(p_q,r)\triangleq\argmin\biggl\{l\in[\ell_q^{(r)}]: \sigma_q^{(r)}(l)\neq m_q(l)\biggr\},\label{eq:prog}
\end{equation}
where recall that $m_q^{\RC(\pi)}$ is the received transcript in the noiseless scenario. In case no $l\in[\ell_q^{(r)}]$ satisfies $\sigma_q^{(r)}(l)\neq m_q(l)$, set $\rp(p_q,r)=\ell_q^{(r)}+1$. We point out here that to measure progress at step $r$, it is enough to compare only the estimate of the received string, $\sigma_q^{(r)}$, with the received transcript in the noiseless setting, $m_q^{\RC(\pi)}$. This is because, by design, Algorithms~\ref{alg:rsperip} and ~\ref{alg:rscent} ensure that the estimated transmitted string, $z_q^{(r)}$, will agree with the (noiseless) sent transcript, $b_q^{\RC(\pi)}$, up to the same length as $\rp(p_q,r)$. This follows from the fact that if a non-$\BK$ symbol is sent by $p_q$ at step $r$, it is decided according to $\sigma_q^{(r)}$ and its input $x_q$.

Next, denote by $\B(p_q,r)$ the number of steps for which $p_q$ has broadcast $\BK$. Note that in Algorithm~\ref{alg:rscent}, in any step $r$, $p_0$ either broadcasts $\BK$ to all $\bp_i,i\in[n_2]$, or to none at all. Thus, using the definition of the PARSE operation in Definition~\ref{def:parse}, any party $p_q$ satisfies 
\begin{equation}
r=\ell_q^{(r)}+1+2\B(p_q,r).\label{eq:back}
\end{equation}

Next, we define the notion of a `historical path.' Firstly, given any two party-step pairs $(p_{q_1},r_1)$ and $(p_{q_2},r_2)$, with $r_2\geq r_1$, we say that $(p_{q_1},r_1)$ lies in the history of $(p_{q_2},r_2)$ if either of the following hold:
\begin{itemize}
\item $p_{q_1}=p_{q_2}$,
\item Either $p_{q_1}=p_0$ or $p_{q_2}=p_0$ and $r_2-r_1\geq 1$,
\item Either $p_{q_1},p_{q_2}\in\bp_i$, for some $i\in[n_2]$, and $r_2-r_1\geq 1$,
\item $r_2-r_1\geq 2$.
\end{itemize}
In other words, we say that $(p_{q_1},r_1)$ lies in the history of $(p_{q_2},r_2)$, if the transmission from $p_{q_1}$ at step $r_1$ will start affecting the transmissions from $p_{q_2}$ within at most $r_2-r_1$ steps. Hence, given any party $p_q$, a party $p_{q'}\neq p_q$ will be referred to as a \emph{neighbour} of $p_q$ if $(p_{q'},r-1)$ lies in the history of $(p_q,r)$. We call a set $\phi=\{(p_{q_{r'}},r'):r'\in[r]\}$ to be a historical path ending at $(p_{q_r},r)$, given $(p_{q_{r'}},r')$ lies in the history of $(p_{q_{r'+1}},r'+1)$, for all $r'\in[r-1]$. 

Next, we state two important technical lemmas. The first lemma relates the the progress of a party at step $r$ with the number of tree code decoding errors affecting its history.

\begin{lemma}
\label{lem:errpath}
Given any party-time pair $(p_q,r)$, there exists a historical path $\phi=\{(p_{q_{r'}},r'):r'\in[r]\}$ with $q_r=q$, such that 
\begin{itemize}
\item There exists $\dfrac{r-\rp(p_q,r)}{2}$ party-time pairs $(p_{q_{r'}},r')\in\phi$, such that $p_{q_{r'}}$ makes at least one tree code-decoding error in step $r'$.
\item There is no decoding error at $(p_{q_{r'}},r')$ if $p_{q_{r'}}\neq p_{q_{r'-1}}$.
\end{itemize}
\end{lemma}

Recall that in the simulation $\Pi$, a tree code decoding takes place after symbols from $S$, the output alphabet of the tree code, are first recovered through the BCast$()$ algorithm (see Algorithms~\ref{alg:rsperip} and~\ref{alg:rscent}). Note that in any step $r$, a party $p_{i,j}$ decodes $n_1$ symbols from $S$ after running BCast$()$. The central party $p_0$ decodes $n_1n_2$ of them. We shall say that a party $p_q$ made a decoding error in BCast$()$ at step $r$, if it decodes at least one of these symbols from $S$ incorrectly. The following lemma relates the progress of a party at step $r$ with the number of decoding errors in BCast$()$ in its history.

\begin{lemma}
\label{lem:ce}
Given any historical path $\phi$ ending at $(p_q,r)$ and satisfying the hypothesis of Lemma~\ref{lem:errpath}, there exists a historical path $\phi'$ also ending at $(p_q,r)$ which satisfies the following: There exists $\dfrac{r-\rp(p_q,r)}{8}$ party-time pairs $(p_{q_{r'}},r')\in\phi'$, such that $p_{q_{r'}}$ makes a decoding error in BCast$()$ at step $r'$.
\end{lemma}
The proofs of these lemmas appear in Appendices~\ref{app:errpath} and \ref{app:ce}.

We now use Lemmas~\ref{lem:errpath} and \ref{lem:ce} to bound the probability of failure of $\Pi$. Consider the event that a certain party $p_q$ has failed to output the correct transcript, i.e., $\rp(p_q,2\RC(\pi)+1)<\RC(\pi)$. Then, by Lemmas~\ref{lem:errpath} and \ref{lem:ce}, there exists a historical path $\phi$ ending at $(p_q, 2\RC(\pi)+1)$ containing at least $\dfrac{\RC(\pi)+1}{8}\geq\dfrac{\RC(\pi)}{16}$ party-step pairs $(p_{q_r},r)$ where party $p_{q_r}$ made a decoding error in BCast$()$ in step $r$. Recall that by the hypothesis of Theorem~\ref{th:rs}, the probability of any party erroneously decoding a symbol from $S$ is $p$. Then, taking union bound over the number of symbols from $S$ decoded during a run of BCast$()$, the probability of $p_{q_r}$ making a decoding error in BCast$()$ at step $r$ is upper bounded by $\zeta=n_1n_2p$. Now, fix any historical path $\phi$ ending in $(p_q,2\RC(\pi)+1)$. Then, the probability that there are more than $\frac{1}{16}\RC(\pi)$ pairs $(p_{q_r},r)$ in $\phi$ where $p_{q_r}$ commits a decoding error in BCast$()$ at step $r$, is upper bounded by
\begin{gather}
\sum_{l=\frac{1}{16}\RC(\pi)}^{2\RC(\pi)+1}{{2\RC(\pi)+1}\choose{l}}\zeta^l(1-\zeta)^{2\RC(\pi)+1-l} \leq 2^{2\RC(\pi)+1}\zeta^{\frac{1}{16}\RC(\pi)}.\label{eq:rs:1}
\end{gather}
Next, note that the total number of historical paths ending at $(p_q,2\RC(\pi)+1)$ can be upper bounded by $(n_1n_2+1)^{2\RC(\pi)}$. This follows by noting that given any pair $(p_{q_{r+1}},r+1)$, the number of pairs $(p_{q_r},r)$ that lie in its history is upper bounded by $n_1n_2+1$.\footnote{The upper bound is achieved with equality if $p_{q_{r+1}}=p_0$, but is loose otherwise.} Thus, by union bound over the total number of historical paths that terminate at $(p_q,2\RC(\pi)+1)$, and by \eqref{eq:rs:1}, the probability that $p_q$ outputs the wrong transcript is upper bounded by 
\begin{gather*}
    (n_1n_2+1)^{2\RC(\pi)}2^{2\RC(\pi)+1}\zeta^{\frac{1}{16}\RC(\pi)}\leq(n_1n_2+1)^{2\RC(\pi)}(n_1n_2)^{\frac{1}{16}\RC(\pi)}2^{2\RC(\pi)+1}p^{\frac{1}{16}\RC(\pi)}\leq n^{4\RC(\pi)}2^{2\RC(\pi)+1}p^{\frac{1}{16}\RC(\pi)}
\end{gather*}
where the first inequality follows by noting $\zeta=n_1n_2p$, and the final one uses $n=n_1n_2+1$. Finally, taking the union bound over all $n$ possible choices for the party $p_q$, the probability of failure of $\Pi$ can be upper bounded by
\begin{gather*}
    n^{4\RC(\pi)+1}2^{2\RC(\pi)+1}p^{\frac{1}{16}\RC(\pi)}\leq (32n^5p^{\frac{1}{16}})^{\RC(\pi)},
\end{gather*} 
as required.

This completes the proof of Theorem~\ref{th:rs} modulo the proofs of Lemmas~\ref{lem:errpath} and \ref{lem:ce} which follow. 

\subsection{Proof of Lemma~\ref{lem:errpath}}\label{app:errpath}

Define the quantity
$$
\X(p_q,r)\triangleq\max_{\phi \text{ ending at }(p_q,r)}|\{(p_{q_{r'}},r')\in\phi: p_{q_{r'}} \text{ makes tree code decoding error at step }r\}|.
$$
We claim the following.
\begin{claim}
For any party $p_q$ and any step $r$, the following holds true:
$$
r\leq \rp(p_q,r)+\B(p_q,r)+\X(p_q,r).
$$
\label{claim:rs}
\end{claim}
\begin{IEEEproof}
We prove the claim using induction on $r$. Notice that since by definition $\rp(p_q,r)\geq 1$, the claim is valid for $r=1$. We now assume the claim is valid for all pairs $(p_{q'},r')$ where $r'\leq r$. We proceed to prove the claim for $(p_q,r+1)$ which by induction will complete the proof of the claim.

To proceed, we classify the actions taken by any party $p_{q'}$ at step $r$ into the following five categories.
\begin{enumerate}
\item \textsf{Progress (p)} - Occurs if $\rp(p_{q'},r)=\rp(p_{q'},r-1)+1$.
\item \textsf{Necessary Backup (bk)} - Occurs if $\BK$ was broadcast, and on parsing $w_{q'}^{r},s_{q'}^{r-1}$, this $\BK$ deletes an incorrect symbol. Observe that $\bk$ implies $\rp(p_{q'},r-1)=\rp(p_{q'},r)$.
\item \textsf{Harmful tree error (hte)} - Occurs if there is a tree code decoding error at $r$, and $\bk$ or $\textsf{p}$ doesn't occur.
\item \textsf{Propagated error (pe)} - Occurs if there is no tree code decoding error at $r$, but $\BK$ is not transmitted and $\rp(p_{q'},r)=\rp(p_{q'},r-1)$.
\item \textsf{Harmful backup (hbk)} - Occurs if there is no tree code decoding error at $r$, $\BK$ is broadcast, and $\rp(p_{q'},r)=\rp(p_{q'},r-1)-1$.
\end{enumerate}
Observe that the above list exhausts all the possible situations.

Now, note that if the action by $p_q$ at $r+1$ is either $\textsf{p}$ or $\bk$, then by the induction hypothesis the claim immediately follows. Assuming $\hte$ occurred, we immediately have $\X(p_q,r+1)=\X(p_q,r)+1$. Therefore, if $\rp(p_q,r+1)=\rp(p_q,r)$ then the claim follows by the induction hypothesis. In case $\rp(p_q,r+1)=\rp(p_q,r)-1$, we must have $\B(p_q,r+1)=\B(p_q,r)+1$, and hence, the claim again follows by induction. We now focus on the two non-trivial cases, $\pe$ and $\hbk$.

First, assume that $\pe$ has occurred, and hence $\B(p_q,r)=\B(p_q,r+1)$. Since, there was a transmission by $p_q$ at $r+1$, we must have the ConsCheck$()$ (Algorithm~\ref{alg:cons}) declare that the transcript decoded by $p_q$ at step $r+1$ is consistent with its estimated transcript. This implies that $\ell_q^{(r)}\leq\hat{\ell}_q^{(r+1)}-1$. This and the fact that $p_q$ made no tree code decoding error at $r+1$, implies that any neighbour $p_{q'}$ of $p_q$ satisfies $\ell_{q'}^{(r)}\geq \ell_{q}^{(r)}$. This immediately implies that $\B(p_{q'},r)\leq \B(p_q,r)$. On the other hand, the fact that ConsCheck$()$ was passed by $p_q$ at $r+1$ and made no tree code decoding errors, the only way in which $\rp(p_q,r)=\rp(p_q,r+1)$ can occur is if there exists some neighbour $p_{q'}$ of $p_q$ satisfying $\rp(p_{q'},r)\leq \rp(p_q,r)-1$. Together with the facts that $\B(p_{q'},r)\leq \B(p_q,r)=\B(p_q,r+1)$, and that $\X(p_{q'},r)\leq\X(p_q,r+1)$ since $p_{q'}$ is a neighbour of $p_q$, the claim follows using the induction hypothesis at $(p_{q'},r)$.

Next, assume that $\hbk$ has occurred, and therefore $\B(p_q,r+1)=\B(p_q,r)+1$ and $\rp(p_q,r+1)=\rp(p_q,r)-1$. Firstly, note that $\rp(p_q,r+1)=\rp(p_q,r)-1$ can occur only if 
\begin{equation}
\rp(p_q,r)=\ell_q^{(r)}+1.\label{eq:claim:1}
\end{equation}. 
Next, the fact that $\BK$ was transmitted by $p_q$ at step $r+1$ implies that it failed the ConsCheck$()$ (Algorithm~\ref{alg:cons}). This can happen in two ways. The first possibility is that $\hat{\ell}_q^{(r+1)}\leq\ell_q^{(r)}$. The second possibility is that the estimated transcript of $p_q$ at step $r+1$ did not agree with the decoded transmissions of other parties, i.e., there exists $l\in[\ell_q^{(r)}]$ such that $\hat{z}_q^{(r+1)}(l)\neq\sigma_q^{(r)}(l)$ or $z_q^{(r)}(l+1)\neq\pi(x_q,\sigma_q^{(r)l})$.

Consider the first scenario, i.e., $\hat{\ell}_q^{(r+1)}\leq\ell_q^{(r)}$. This implies that there exists some neighbour $p_{q'}$ of $p_q$ satisfying $\ell_{q'}^{(r)}+1\leq\ell_q^{(r)}$. Noting that $\ell_{q'}^{(r)}$ and $\ell_q^{(r)}$ are obtained by parsing strings of the same length, both lengths shall either be even or odd, and hence, this further implies that $\ell_{q'}^{(r)}+2\leq\ell_{q}^{(r)}$. This in turn means that $\B(p_{q'},r)\geq \B(p_q,r)+1$. Finally, recalling \eqref{eq:back} and \eqref{eq:claim:1}, we have
\begin{align*}
\rp(p_q,r) & = r-2\B(p_q,r)\\
           & \stackrel{(a)}{\geq} \rp(p_{q'},r)+2(\B(p_{q'},r)-\B(p_q,r))\\
           & \stackrel{(b)}{\geq}\rp(p_{q'},r)+\B(p_{q'},r)-\B(p_q,r)+1,
\end{align*}
and thus,
\begin{equation}
\rp(p_q,r)+\B(p_q,r)-1\geq\rp(p_{q'},r)+\B(p_{q'},r),\label{eq:claim:2}
\end{equation}
where (a) follows using \eqref{eq:back} and (b) made use of the fact that $\B(p_{q'},r)\geq \B(p_q,r)+1$

Now, in the second scenario, assume $\hat{\ell}_q^{(r+1)}>\ell_q^{(r)}$, but the decoded transmission did not agree with the estimated transcript. Firstly, $\hat{\ell}_q^{(r+1)}>\ell_q^{(r)}$ implies that all neighbours $p_{q'}$ of $p_q$ satisfy $\ell_{q'}^{(r)}\geq\ell_q^{(r)}$ and hence $\B(p_{q'},r)\leq \B(p_q,r)$. Secondly, note that since $\rp(p_q,r)=\ell_q^{(r)}+1$, and it made no tree code decoding error, the only way the estimated transcript can be inconsistent if there exists a neighbour $p_{q'}$ satisfying $\rp(p_{q'},r)\leq\rp(p_q,r)-1$. Therefore, we obtain,
\begin{equation}
\rp(p_q,r)+\B(p_q,r)-1\geq\rp(p_{q'},r)+\B(p_{q'},r).\label{eq:claim:3}
\end{equation}
Combining \eqref{eq:claim:2} and \eqref{eq:claim:3}, we have that $\hbk$ occurring implies that there exists a neighbour $p_{q'}$ satisfying $\rp(p_q,r)+\B(p_q,r)-1\geq\rp(p_{q'},r)+\B(p_{q'},r)$. Moreover, by noting that $\hbk$ implies $\rp(p_q,r)+\B(p_q,r)=\rp(p_q,r+1)+\B(p_q,r+1)$, and that $\X(p_q,r+1)\geq\X(p_{q'},r)$ since $p_{q'}$ is a neighbour of $p_q$, the claim follows using the induction hypothesis at $(p_{q'},r)$.
\end{IEEEproof}

One important consequence of the proof of Claim~\ref{claim:rs} is as follows. For any point $(p_q,r)$, there exists a historical path $\phi$ ending at $(p_q,r)$ that satisfies the following:
\begin{itemize}
\item There is no decoding error at $(p_{q_{r'}},r')$ if $p_{q_{r'}}\neq p_{q_{r'-1}}$.
\item The total number of pairs $(p_{q_{r'}},r')\in\phi$ where $p_{q_{r'}}$ makes a tree code decoding error at step $r'$, denoted by $\E(\phi)$ satisfies $r\leq \rp(p_q,r)+\B(p_q,r)+\E(\phi)$.
\end{itemize}
This holds because, in the induction proof of the claim, the only actions for which we need to extend the historical path from $(p_{q'},r)$ to $(p_q,r+1)$ (with $p_q\neq p_{q'}$), in order to maintain the hypothesis of the claim, are $\pe$ and $\hbk$. It turns, that in both of these cases there are no tree code decoding errors at $(p_q,r+1)$, and hence, the historical path $\phi$ extended in this manner satisfies the properties mentioned above.

To complete the proof of Lemma~\ref{lem:errpath}, it remains to show that $\E(\phi)\geq \displaystyle\frac{r-\rp(p_q,r)}{2}$. This is true, since
\begin{align*}
\E(\phi) & \geq \frac{r+r-2\B(p_q,r)-2\rp(p_q,r)}{2}\\
         & \stackrel{(a)}{\geq}\frac{r-\rp(p_q,r)}{2},
\end{align*}
where (a) uses \eqref{eq:prog} and \eqref{eq:back}. 

\subsection{Proof of Lemma~\ref{lem:ce}}\label{app:ce}

To begin with, we make the following observation. For any pair of strings $y_1^l$, $y_2^l$, we call $h$ the length of their suffix mismatch, if the length of their longest prefix is $l-h$. Now, recall that any pair of valid codewords $y_1^l,y_2^l\in S^l$, encoded using the tree code $\cT$ with distance $\alpha=\frac{1}{2}$, must satisfy $\Delta(y_1^l,y_2^l)\geq \frac{h}{2}$, where $h$ is the length of their suffix mismatch. Thus, noting the decoding rule in \eqref{eq:dec} used by the parties, a party $p_q$, while decoding the transmissions from its neighbour $p_{q'}$, can make a tree-code decoding error of suffix mismatch $h$, only if it has made at least $\frac{h}{4}$ errors in decoding the symbols from $S$ sent by $p_{q'}$ in the last $h$ runs of BCast$()$. 

Next, consider the historical path $\phi$ ending in $(p_q,r)$ given by Lemma~\ref{lem:errpath}. We construct the required historical path $\phi'$ ending at $(p_q,r)$ as follows. Begin with $(p_q,r)$ and trace $\phi$ backwards till a point $(p_{q'},r')$ is obtained which corresponds to a tree code decoding error. Let $h$ be the length of the maximal suffix mismatch of this decoding error.\footnote{Note that $p_{q'}$ decodes transmissions from all its neighbours, and it need not be that it makes an error in decoding only one of these transmissions. We consider only the decoding with the largest suffix mismatch.} For the next $h-1$ steps $\phi'$ remains at $p_{q'}$. This implies that $p_{q'}$ makes an error is decoding BCast$()$ in at least $\frac{h}{4}$ steps between $r'-h+1$ to $r'$. In other words, $\phi'$ contains at least $\frac{h}{4}$ steps between $r'-h+1$ and $r$ with decoding errors in BCast$()$. Now, consider the following cases.
\begin{enumerate}
    \item If $(p_{q'},r-h+1)\in\phi$ we extend $\phi'$ by continuing to trace $\phi$ again.
    \item If $p_{q'}=p_0$, then $\phi'$ joins $\phi$ in step $r-h$.
    \item If $(p_0,r'-h)\in\phi$, then also $\phi'$ joins $\phi$ in step $r-h$.
    \item If the party associated with the $(r'-h)$th step of $\phi$ belongs to the same broadcast link as $p_{q'}$, then also $\phi'$ joins $\phi$ in step $r-h$.
    \item If $p_{q'}\neq p_0$, and the party associated with step $r'-h$ in $\phi$ is not from the same broadcast link as $p_q$, then $\phi'$ joins $\phi$ in the $(r'-h-1)$th step as follows. Include $(p_0,r'-h)\in\phi'$. Then obviously $\phi'$ can rejoin $\phi$ in the next step.
\end{enumerate}
After rejoining $\phi$, $\phi'$ continues to track it till it encounters the next tree code error, and then follows the same procedure. Next we bound the number of locations with tree code error in $\phi$ between (including) the steps $r'$ and the last step before $\phi'$ joins $\phi$. In the first four cases 1)-4), this can be at most $h$ since there are only $h$ such steps. In case 5) there are $h+1$ steps. However, there is at least one step in which the party associated with it changes from the previous step, as otherwise we will have case 1) and not case 5). By the hypothesis of Lemma~\ref{lem:errpath}, this step cannot have a tree code error. Thus even in this case, we will have at most $h$ tree code errors. This allows us to conclude that $\phi'$ has at least $\frac{l}{4}$ steps with decoding error in BCast$()$, given $\phi$ has $l$ steps with tree code decoding errors. Combining this with the fact that $\phi$ has $\dfrac{r-\rp(p_q,r)}{2}$ steps with tree code decoding errors, as given by Lemma~\ref{lem:errpath}, the path $\phi'$ will have $\dfrac{r-\rp(p_q,r)}{8}$ steps with decoding errors in BCast$()$ as required.

\section{Proof of Theorem~\ref{th:star}}\label{app:star}

Throughout this proof, we assume that $\RC(\pi)=\omega(k)$. If not, i.e, if $\RC(\pi)=O(k)$, we claim that a very simple simulation $\Pi$ (not the one in Algorithm~\ref{alg:star}) will give us the result of Theorem~\ref{th:main}. We give a rough sketch below. Firstly, if $\RC(\pi)=O(k)$, then the number of steps in $\tilde{\pi}$ is a constant, since $\lceil\frac{1}{k}\RC(\pi)\rceil=O(1)$. Then, instead of an elaborate simulation, we can simply protect all transmissions of $\tilde{\pi}$ by simple error correcting codes. Note that the alphabet size of the symbols sent by $p_{i,j}$ and $p_0$ in $\tilde{\pi}$ are respectively $\Theta(2^{n_1k})=\Theta(\log n_2)$ and $\Theta(k)$. Hence, similar to Lemma~\ref{lem:eccex}, these transmissions can be encoded using error correcting codes of output length $O(\log n_2)$, which allow correct decoding except with probability $n_2^{-3}$ in presence of $\epsilon$-bit flip noise. Now, note that in any step, a party $p_{i,j}$ decodes $n_1$ transmissions, while $p_0$ decodes $n_1n_2$ transmissions. Thus, the total number of decodings per step is $n_1^2n_2+n_1n_2$, which can be  upper bounded by $n_2^2$, since in this regime $n_2>2^{n_1}$. Noting, that we have only $O(1)$ steps, and each step has at most $n_2^2$ decodings, each with a failure probability of $n_2^{-3}$, a union bound over all possible decodings show that the simulation is successful, except with probability $O(1)n_2^2n_2^{-3}=n_2^{-\Omega(1)}$. Furthermore, note that each step has two stages, both involving broadcasting error correcting codes of output length $O(\log n_2)$. In other words, each step costs $O(\log n_2)$ rounds. Thus, the round complexity of this simple simulation is $2\lceil\frac{1}{k}\RC(\pi)\rceil O(\log n_2)=\RC(\pi)O(\frac{n_1\log n_2}{\log\log n_2})$ as required. Hence, we shall drop this case, and consider $\RC(\pi)=\omega(k)$ while establishing the failure probability of the simulation $\Pi$ given in Algorithm~\ref{alg:star}.

The proof proceeds similar to that of Theorem~\ref{th:rs} in Appendix~\ref{app:rs}, and is sketched below. First, recall that $m_q^{\RC(\pi)}$ is the correct received transcript of the party $p_q$ in absence of noise. Thus, define the real progress of the parties $p_q$ after $r$ steps as follows. 
\begin{equation}
\rp(p_q,r)\triangleq
\begin{cases}
\argmin\biggl\{l\in[\ell_q^{(r)}]: \cpath_q(\sigma_q^{(r)l})\neq m_q^l\biggr\}-1, & \text{ if }q=0,\\
\argmin\biggl\{l\in[\ell_q^{(r)}]: \cpath_q(\sigma_q^{(r)l})\neq m_q^l\biggr\}, & \text{ otherwise}.\label{eq:progstar}
\end{cases}
\end{equation}
In case no $l\in[\ell_q^{(r)}]$ satisfies $\cpath_q(\sigma_q^{(r)l})\neq m_q^l$, set $\rp(p_q,r)=\ell_q^{(r)}$, if $q=0$, and $\rp(p_q,r)=\ell_q^{(r)}+1$, otherwise. Similarly, denote by $\B(p_q,r)$ the number of steps $p_q$ has broadcast $\BK$, and hence, any party $p_q$ satisfies 
\begin{equation}
r=
\begin{cases}
\ell_q^{(r)}+2\B(p_q,r), & \text{ if }q=0,\\
\ell_q^{(r)}+1+2\B(p_q,r), & \text{ otherwise}.\label{eq:backstar}
\end{cases}
\end{equation}

Next, we will define the notion of the historical path in a slightly different way. A historical path $\phi$ ending at $(p_0,r)$ is any set $\biggl\{(p_{q_{r'}},r'):r'\in[r]\biggr\}\bigcup\biggl\{(p_0,r'):r'\in[r]\biggr\}$ where $p_{q_{r'}}\in\bigcup_{i\in[n_2]}\bp_i$, for all $r'\in[r]$. On the other hand, for some party $p_q\neq p_0$, a historical path $\phi$ ending at $(p_q,r)$ is any set $\biggl\{(p_{q_{r'}},r'):r'\in[r]\biggr\}\bigcup\biggl\{(p_0,r'):r'\in[r-1]\biggr\}$ where $p_{q_r}=p_q$ and $p_{q_{r'}}\in\bigcup_{i\in[n_2]}\bp_i$, for all $r'\in[r-1]$. We say that a location $(p_{q_{r'}},r')\in\phi$ makes an error if party $p_{q_{r'}}$ makes a decoding error in at step $r'$. Then, we have the following lemmas. 
\begin{lemma}
\label{lem:errpathstar}
Given any party-time pair $(p_q,r)$, there exists a historical path $\phi$ ending at $(p_q,r)$ containing at least $\dfrac{r-\rp(p_q,r)}{2}$ locations with a tree code decoding error.
\end{lemma}

\begin{lemma}
\label{lem:cestar}
There exists a historical path $\phi'$ ending at $(p_q,r)$ containing at least $\dfrac{r-\rp(p_q,r)}{8}$ locations with an error correcting code decoding error. 
\end{lemma}
The proof of Lemmas~\ref{lem:errpathstar} and \ref{lem:cestar} are similar to that of Lemmas~\ref{lem:errpath} and \ref{lem:ce}, and are sketched in Appendices~\ref{app:errpathstar} and \ref{app:cestar} respectively.

We now use Lemmas~\ref{lem:errpathstar} and \ref{lem:cestar} to bound the probability of failure of $\Pi$. Firstly, recall that in this regime $n_2=2^{2^{\Omega(\frac{n_1\log n_1}{\log\log n_1})}}$. Hence, we have $n_1n_2\leq n\leq n_2^2$. Define $R\triangleq2\lceil\frac{1}{k}\RC(\pi\rceil$. Consider the event that a certain party $p_q$ has failed to output the correct transcript, i.e., $\rp(p_q,R)<\lceil\frac{1}{k}\RC(\pi)\rceil$, if $q=0$, or $\rp(p_q,R+1)\leq\lceil\frac{1}{k}\RC(\pi)\rceil$, otherwise. Then, by Lemma~\ref{lem:cestar}, there exists a historical path $\phi$ ending at $(p_q,R)$ if $q=0$, or $(p_q,R+1)$, otherwise, such that $\phi$ contains at least $\frac{1}{8k}\RC(\pi)$ locations with an error correcting code decoding error. By Lemma~\ref{lem:eccex}, a party can make an error while decoding the error correcting coded transmission of another party with probability $p\triangleq n_2^{-52}$. Further, at any given step, a party $p_{q'}$ will decode a maximum of $n_1n_2\leq n$ error correcting coded symbols.\footnote{Note here that the maximum number of decoding is done by the central party $p_0$ which decodes $n_1n_2$ symbols coming from all the parties in $\bigcup_{i\in[n_2]}\bp_i$.} Thus, the probability that a location $(p_{q_{r'}},{r'})$ contains an error correcting code decoding error is upper bounded by $\zeta=np$. Now, fix any historical path $\phi$ ending in $(p_q,R)$ if $q=0$, or $(p_q, R+1)$ otherwise. Then, the probability that at least $\frac{1}{8k}\RC(\pi)$ locations in $\phi$ make an error correcting code decoding error is upper bounded by
\begin{gather}
\sum_{l=\frac{1}{8k}\RC(\pi)}^{R+1}{{R+1}\choose{l}}\zeta^l\leq 2^{R+1}\zeta^{\frac{1}{8k}\RC(\pi)}.\label{eq:star:1}
\end{gather}
Next, note that the total number of historical paths ending at $(p_q,R)$, if $q=0$, or $(p_q,R+1)$, otherwise, can be upper bounded by $(n_1n_2)^{R}\leq n^R$. Therefore, by union bound over the total number of historical paths and using \eqref{eq:star:1}, the probability that $p_q$ outputs the wrong transcript is upper bounded by $n^R2^{R+1}(np)^{\frac{1}{8k}\RC(\pi)}$. Taking the union bound over all $n$ possible choices for the party $p_q$, the probability of failure of $\Pi$ is thus upper bounded by
\begin{gather*}
    n^{R+1}2^{R+1}(np)^{\frac{1}{8k}\RC(\pi)}\stackrel{(a)}{\leq}(8n^3n^{\frac{1}{8}}p^{\frac{1}{8}})^{\frac{1}{k}\RC(\pi)}\stackrel{(b)}{\leq} (8n^{\frac{25}{8}}n_2^{-\frac{52}{8}})^{\frac{1}{k}\RC(\pi)}\stackrel{(c)}{\leq}(8n^{-\frac{1}{8}})^{\frac{1}{k}\RC(\pi)}\leq n^{-\Omega(\frac{1}{k}\RC(\pi))},
\end{gather*}
which vanishes since $\RC(\pi)=\omega(k)$. Here, $(a)$ follows by noting $R=2\lceil\frac{1}{k}\RC(\pi)\rceil$, $(b)$ is obtained by plugging in $p\leq n_2^{-52}$, and $(c)$ uses the fact that in this regime $n\leq n_2^2$. This completes the proof of Theorem~\ref{th:star} modulo the proofs of Lemmas~\ref{lem:errpathstar} and \ref{lem:cestar} which follow. 

\subsection{Proof of Lemma~\ref{lem:errpathstar}}\label{app:errpathstar}

Define the quantity
$$
\X(p_q,r)\triangleq\max_{\phi \text{ ending at }(p_q,r)}|\{(p_{q_{r'}},r')\in\phi: p_{q_{r'}} \text{ makes tree code decoding error at step }r\}|.
$$
We claim the following.
\begin{claim}
For any party $p_q$ and any step $r$, the following holds true:
\begin{equation*}
r\leq \rp(p_q,r)+\B(p_q,r)+\X(p_q,r)
\end{equation*}
\label{claim:rsstar}
\end{claim}
\begin{IEEEproof}
The claim is proved by induction on $r$. The base case $r=1$ is immediately true for all parties $p_q\neq p_0$, since by definition $\rp(p_q,1)\geq 1$. For $p_0$, note that either it makes a decoding error in the first step, or it sends the correct information. In other words, either $X(p_0,1)=1$ or $\rp(p_0,1)=1$. Thus the base case for $r=1$ is true. Assuming the claim is valid for all pairs $(p_{q'},r')$ where $r'\leq r$, we proceed to prove the claim for $(p_q,r+1)$.

To proceed, recall the five categories in which the action taken by any party at step $r$ can be classified.
\begin{enumerate}
\item \textsf{Progress (p)} - Occurs if $\rp(p_{q'},r)=\rp(p_{q'},r-1)+1$.
\item \textsf{Necessary Backup (bk)} - Occurs if $\BK$ was broadcast, and on parsing $w_{q'}^r,s_{q'}^{r-1}$ (or $s_{q'}^r$ if $q'=0$), this $\BK$ deletes an incorrect symbol. Observe that $\bk$ implies $\rp(p_{q'},r-1)=\rp(p_{q'},r)$.
\item \textsf{Harmful tree error (hte)} - Occurs if there is a tree code decoding error at $r$, and $\bk$ or $\textsf{p}$ doesn't occur.
\item \textsf{Propagated error (pe)} - Occurs if there is no tree code decoding error at $r$, but $\BK$ is not transmitted and $\rp(p_{q'},r)=\rp(p_{q'},r-1)$.
\item \textsf{Harmful backup (hbk)} - Occurs if there is no tree code decoding error at $r$, $\BK$ is broadcast, and $\rp(p_{q'},r)=\rp(p_{q'},r-1)-1$.
\end{enumerate}
Similar to the proof of Lemma~\ref{lem:errpath}, it is easy to show the validity of the induction hypothesis at $(p_q, r+1)$ if \textsf{p}, \textsf{bk}, or \textsf{hte} has occurred. We therefore directly proceed to prove the result for the cases \textsf{pe} and \textsf{hbk}.

First, assume that $\pe$ has occurred, and hence $\B(p_q,r)=\B(p_q,r+1)$. Since, there was a transmission by $p_q$ at $r+1$, we must have the ConsCheckChunked$()$ (Algorithm~\ref{alg:consstar}) declare that the transcript decoded by $p_q$ at step $r+1$ is consistent with its estimated transcript. This implies that $\ell_q^{(r)}\leq\hat{\ell}_q^{(r+1)}-1$. We shall first prove that the induction hypothesis is valid for $(p_{i,j},r+1)$ for all $i\in[n_2], j\in[n_1]$. We shall then exploit this to show the validity for $(p_0,r+1)$. Noting that there was no tree code decoding error, $\ell_{i,j}^{(r)}\leq\hat{\ell}_{i,j}^{(r+1)}-1$, implies that $\ell_{i,j}^{(r)}\leq\ell_{i,j'}^{(r)}$, for all $j'\neq j$, as well as $\ell_{i,j}^{(r)}\leq \ell_0^{(r)}-1$. Hence, we have $\B(p_{i,j},r)\geq\B(p_{i,j'},r), j'\neq j$, as well as $\B(p_{i,j},r)\geq\B(p_0,r)$. Then, noting that $p_{i,j}$ sent a wrong symbol in round $r+1$, there must exist at least one party $p_{q'}\neq p_{i,j}$ such that $\rp(p_{i,j},r)\geq \rp(p_{q'},r)+1$. Together with the facts that $\B(p_{q'},r)\leq \B(p_{i,j},r)=\B(p_{i,j},r+1)$, $\rp(p_{i,j},r)=\rp(p_{i,j},r+1)$, and that $\X(p_{q'},r)\leq\X(p_{i,j},r+1)$, the claim follows using the induction hypothesis at $(p_{q'},r)$.

Next, for $p_0$, the condition $\ell_q^{(r)}\leq\hat{\ell}_q^{(r+1)}-1$ translates to $\ell_0^{(r)}\leq\ell_{i,j}^{(r+1)}$ since there was no decoding error,\footnote{Note that we have $\ell_{i,j}^{(r+1)}$ instead of $\ell_{i,j}^{(r)}$ since $p_0$ decodes after the completion of the first part, and hence it is decoding an $r+1$ length string.} and hence $\B(p_0,r)\geq \B(p_{i,j},r+1)$ for all $i\in[n_2], j\in[n_1]$.\footnote{Again observe that we have $\B(p_{i,j},r+1)$ instead of $\B(p_{i,j},r)$ since the operations at $p_0$ take place only after the operations at $p_{i,j}$ for the $(r+1)$th step has been completed.} Moreover, the fact that $p_0$ transmitted a wrong symbol implies that there exists at least one party $p_{q'}\neq p_0$ satisfying $\rp(p_{q'},r+1)\leq\rp(p_0,r)$. Together with the fact that $\B(p_0,r+1)=\B(p_0,r)\geq \B(p_{q'},r+1)$, $\rp(p_0,r+1)=\rp(p_0,r)$, and $\X(p_0,r+1)\geq \X(p_{q'},r+1)$, the induction hypothesis on $(p_{q'},r+1)$ gives the result.

Finally, consider that $\hbk$ has occurred, and therefore $\B(p_q,r+1)=\B(p_q,r)+1$ and $\rp(p_q,r+1)=\rp(p_q,r)-1$, i.e., $\rp(p_q,r)+\B(p_q,r)=\rp(p_q,r+1)=\B(p_q,r+1)$. Also, note that $\rp(p_q,r+1)=\rp(p_q,r)-1$ can occur only if 
\begin{equation}
\rp(p_q,r)=
\begin{cases}
\ell_q^{(r)} & \text{ if }q=0,\\
\ell_q^{(r)}+1 & \text{ otherwise}.\label{eq:claimstar:1}
\end{cases}
\end{equation}
Again, we shall first prove the induction hypothesis for $(p_{i,j},r+1), i\in[n_2], j\in[n_1]$, and then exploit it to prove the induction hypothesis for $(p_0,r+1)$. Noting that $\BK$ was transmitted by $p_{i,j}$ at step $r+1$ implies that it failed the ConsCheckChunked$()$ (Algorithm~\ref{alg:consstar}). This can happen in two ways. The first possibility is that $\hat{\ell}_{i,j}^{(r+1)}\leq\ell_{i,j}^{(r)}$. The second possibility is that the estimated transcript of $p_{i,j}$ at step $r+1$ did not agree with the decoded transmissions of other parties.

Consider the first scenario, i.e., $\hat{l}_{i,j}^{(r+1)}\leq\ell_{i,j}^{(r)}$. Noting that there was no tree code decoding error, this implies either $\ell_0^{(r)}\leq\ell_{i,j}^{(r)}$, or there exists some party $p_{i,j'}$ for which $\ell_{i,j'}^{(r)}+1\leq\ell_{i,j}^{(r)}$. 

\emph{Case I. Assume $\ell_0^{(r)}\leq\ell_{i,j}^{(r)}$}. Observe that $\ell_0^{(r)}$ is the length of the parsed version of a length $r$ string, whereas $\ell_{i,j}^{(r)}+1$ is the length of the parsed version of a length $r$ string. Thus, noting that $\ell_0^{(r)}$ and $\ell_{i,j}^{(r)}+1$ must both be either odd or even, we must have $\ell_0^{(r)}+1\leq\ell_{i,j}^{(r)}$. This immediately translates to $\B(p_0,r)\geq \B(p_{i,j},r)+1$. Thus, using \eqref{eq:claimstar:1}, we have
\begin{align*}
\rp(p_{i,j},r) & =\ell_{i,j}^{(r)}+1\\
               & = r-2\B(p_{i,j},r)\\
               & \stackrel{(a)}{\geq}\rp(p_0,r)+2\B(p_0,r)-2\B(p_{i,j},r),          
\end{align*}
where $(a)$ follows from \eqref{eq:backstar} by noting $\rp(p_0,r)\leq\ell_0^{(r)}$. Rearranging the terms, and using $\B(p_0,r)\geq \B(p_{i,j},r)+1$, we have $\rp(p_{i,j},r)+\B(p_{i,j},r)\geq\rp(p_0,r)+\B(p_0,r)+1$. Next, recalling that $\rp(p_{i,j},r+1)+\B(p_{i,j},r+1)=\rp(p_{i,j},r)+\B(p_{i,j},r)$ since \textsf{hbk} occurred, we have $\rp(p_{i,j},r+1)+\B(p_{i,j},r+1)\geq\rp(p_0,r)+\B(p_0,r)+1$. Finally, noting that $\X(p_{i,j},r+1)\geq\X(p_0,r)$, the induction hypothesis on $(p_0,r)$ gives the result.

\emph{Case II. Assume there exist $p_{i,j'}$ satisfying $\ell_{i,j'}^{(r)}+1\leq\ell_{i,j}^{(r)}$}. Here again, noting that $\ell_{i,j'}+1$ and $\ell_{i,j}+1$ are both lengths of parsed versions of $r$-length strings, one can argue, as in the previous case, that $\ell_{i,j'}^{(r)}+2\leq\ell_{i,j}^{(r)}$ holds. Therefore, we have $\B(p_{i,j'},r)\geq \B(p_{i,j},r)+1$. The rest of the argument follows exactly similar to the previous case and is omitted. 

This completes the proof of for the first scenario that $\hat{l}_{i,j}^{(r+1)}\leq\ell_{i,j}^{(r)}$. Next, consider the remaining scenario where $\hat{l}_{i,j}^{(r+1)}\geq\ell_{i,j}^{(r)}+1$, but $p_{i,j}$ still did not pass the consistency check. Firstly, note that $\hat{l}_{i,j}^{(r+1)}\geq\ell_{i,j}^{(r)}+1$ and no decoding error implies $\ell_0^{(r)}\geq\ell_{i,j}^{(r)}+1$ and $\ell_{i,j'}^{(r)}\geq\ell_{i,j}^{(r)}$ for all $j'\neq j$. This immediately implies that $\B(p_{i,j},r)\geq\B(p_{q'},r)$ for all other $p_{q'}\in\bp_{i+0}$. Secondly, the fact that $p_{i,j}$ still failed the consistency check and there was no decoding error implies that there exists at least one party $p_{q'}\in\bp_{i+0}$ with lesser progress, i.e., satisfying $\rp(p_{i,j},r)-1\geq\rp(p_{q'},r)$. Noting that $\B(p_{i,j},r)\geq\B(p_{q'},r)$, and recalling $\rp(p_{i,j},r+1)+\B(p_{i,j},r+1)=\rp(p_{i,j},r)+\B(p_{i,j},r)$ since \textsf{hbk} occurred, we have $\rp(p_{i,j},r+1)+\B(p_{i,j},r+1)\geq \rp(p_{q'},r)+\B(p_{q'},r)+1$. The proof then follows by noting the induction hypothesis on $(p_{q'},r)$ and $\X(p_{i,j},r+1)\geq\X(p_{q'},r)$.

To complete the proof, it remains to show the validity of the induction hypothesis for $(p_0,r+1)$ given \textsf{hbk} has occurred. The proof is similar to the proof for the case of $(p_{i,j},r+1)$. Again consider the first scenario where $\hat{l}_0^{(r+1)}\leq\ell_0^{(r)}$. No decoding error implies that there exists some party $p_{q'}\neq p_0$ satisfying $\ell_{q'}^{(r+1)}+1\leq\ell_0^{(r)}$. Noting that $\ell_{q'}^{(r+1)}+1$ is the length of the parsed version of an $(r+1)$-length string, whereas $\ell_0^{(r)}$ is the length of the parsed version of an $r$-length string, one can argue that $\ell_{q'}^{(r+1)}+2\leq\ell_0^{(r)}$ holds. This immediately implies $\B(p_{q'},r+1)\geq\B(p_0,r)+1$. Now, recalling that \textsf{hbk} implies $\rp(p_0,r)=\ell_0^{(r)}=r-2\B(p_0,r)$, and noting that $r+1\geq\rp(p_{q'},r+1)+2\B(p_{q'},r+1)$, we have $\rp(p_0,r)\geq \rp(p_{q'},r+1)+2\B(p_{q'},r+1)-2\B(p_0,r)-1$. Rearranging the terms and recalling $\B(p_{q'},r+1)\geq\B(p_0,r)+1$, we get $\rp(p_0,r)+\B(p_0,r)\geq\rp(p_{q'},r+1)+\B(p_{q'},r+1)$. The result then follows by noting $\rp(p_0,r+1)+\B(p_0,r+1)=\rp(p_0,r)+\B(p_0,r)$ as \textsf{hbk} occurred, $\X(p_0,r+1)\geq\X(p_{q'},r+1)$, and the induction hypothesis on $(p_{q'},r+1)$.

It remains to consider the scenario when $\hat{\ell}_0^{(r+1)}\geq\ell_0^{(r)}+1$, but $p_0$ still fails the consistency check for the $(r+1)$th step. Firstly, no decoding error and $\hat{\ell}_0^{(r+1)}\geq\ell_0^{(r)}+1$ shows that $\ell_{q'}^{(r+1)}+1\geq\ell_0^{(r)}+1$, for all parties $p_{q'}\neq p_0$, and hence, $\B(p_0,r)\geq\B(p_{q'},r+1)$. Secondly, the failure of the consistency check despite no decoding error implies that at least one party $p_{q'}\neq p_0$ has made less progress, i.e., $\rp(p_0,r)\geq\rp(p_{q'},r+1)$. Combining these two facts, we get that for some party $p_{q'}\neq p_0$, $\rp(p_0,r)+\B(p_0,r)\geq\rp(p_{q'},r+1)+\B(p_{q'},r+1)$. The proof thus follows noting $\rp(p_0,,r+1)+\B(p_0,r+1)=\rp(p_0,r)+\B(p_0,r)$ as \textsf{hbk} occurred, $\X(p_0,r+1)\geq\X(p_{q'},r+1)$, and the induction hypothesis on $(p_{q'},r+1)$. This completes the proof of the claim.
\end{IEEEproof}

With the claim at hand, note that for any $(p_q,r)$, we have
\begin{align*}
\X(p_q,r) & \geq \frac{r+r-2\B(p_q,r)-2\rp(p_q,r)}{2}\\
          & \stackrel{(a)}{\geq}\frac{r-\rp(p_q,r)}{2},
\end{align*}
where (a) uses \eqref{eq:progstar} and \eqref{eq:backstar}. This implies the existence of a historical path $\phi$ ending at $(p_q,r)$ with tree code decoding error in at least $\frac{r-\rp(p_q,r)}{2}$ locations. This completes the proof of Lemma~\ref{lem:errpathstar}.

\subsection{Proof of Lemma~\ref{lem:cestar}}\label{app:cestar} 

Recall that for a pair of strings $y_1^l$, $y_2^l$, their suffix mismatch is $h$ if the length of their longest prefix is $l-h$. Since both tree codes $\cT_\cc$ and $\cT_\nc$ have distance $\alpha=0.5$, any pair of valid codewords $y_1^l,y_2^l$ must satisfy $\Delta(y_1^l,y_2^l)\geq \frac{h}{2}$, where $h$ is the length of their suffix mismatch. Therefore, recalling that we use minimum distance decoding (see \eqref{eq:dec}) to decode the tree codes, a tree code decoding error of suffix mismatch $h$ must involve at least $\frac{h}{4}$ error correcting code decoding errors in the last $h$ locations. 

Now, consider the historical path $\phi$ given by Lemma~\ref{lem:errpathstar}. We shall construct the required historical path $\phi'$ whose number of steps with error correcting code decoding error is at least $\frac{1}{4}$th as many as the number of steps in $\phi$ with tree code decoding errors. This new historical path $\phi'$, by Lemma~\ref{lem:errpathstar}, will therefore contain $\dfrac{r-\rp(p_q,r)}{8}$ steps with error correcting code decoding error as required. Firstly, consider the locations in $\phi$ with parties $p_q\neq p_0$. Starting from step $r$, $\phi'$ traces $\phi$ backwards until a location $(p_{q'},r')$ is obtained with a tree code decoding error. Let $h$ be the length of the maximal suffix mismatch. Then, add $\{(p_{q'},r'-r''):r''\in[h-1]\}$ to $\phi'$. Noting that the maximal suffix mismatch was $h$, we must have that $p_{q'}$ made error correcting code decoding errors in at least $\frac{h}{4}$ steps between $r'$ and $r'-h+1$. In other words, $\phi'$ has at least $\frac{h}{4}$ steps with error correcting code decoding errors between steps $r'$ to $r'-h+1$. During these steps, the locations in $\phi$ could have made at most $h$ tree code decoding errors, since we only have a total of $h$ steps. We let $\phi'$ join $\phi$ in step $r-h$, and continue to trace it backwards till the next location with tree code decoding error is encountered. This ensures that if $\phi$ has $l$ steps where some non central party makes a tree code decoding error, then $\phi'$ has at least $\frac{l}{4}$ steps where a non central party makes an error correcting code decoding error.

Finally, note that any historical path will contain $(p_0, r')$ for all $r'\in[r]$. Now, starting from $r$ trace back the steps to find the first step $r'$ where $(p_0,r')$ makes a tree code decoding error. Let $h$ be the length of the maximal suffix mismatch. Then, we must have that $p_0$ made error correcting code decoding errors in at least $\frac{h}{4}$ many steps between $r'$ and $r'-h+1$. Continuing this procedure from step $r'-h$ till all $r$ steps have been covered, we can claim that if $p_0$ makes a $l$ tree code decoding errors in a total of $r$ steps, then it must have made at least $\frac{l}{4}$ error correcting code decoding errors in those $r$ steps. Thus $\phi'$ has the required number of steps with error correcting code decoding errors.
\end{appendices}
\end{document}